\documentclass[aps,prx,reprint,superscriptaddress,longbibliography]{revtex4-2}

\usepackage{jabbrv}

\usepackage{amsmath,amsfonts,amsthm,amssymb,bbm,graphicx,color,enumerate,hyperref,xcolor,tikz,subfigure,braket,relsize,lineno}
\usepackage{tabularray, textcomp, multirow, float}
\usepackage[mathscr]{euscript}

\UseRawInputEncoding
\begin{document}

\title{Photo-induced charge carrier dynamics in a semiconductor-based ion trap investigated via motion-sensitive qubit transitions}

%% -- Authors and affiliations
\author{Woojun Lee}
\thanks{These authors contributed equally to this work.}
\affiliation{Dept. of Computer Science and Engineering, Seoul National University, Seoul 08826, South Korea}
\affiliation{Automation and Systems Research Institute, Seoul National University, Seoul 08826, South Korea}
\affiliation{Institute of Computer Technology, Seoul National University, Seoul 08826, South Korea}
\author{Daun Chung}
\thanks{These authors contributed equally to this work.}
\affiliation{Dept. of Computer Science and Engineering, Seoul National University, Seoul 08826, South Korea}
\affiliation{Automation and Systems Research Institute, Seoul National University, Seoul 08826, South Korea}
\author{Honggi Jeon}
\affiliation{Dept. of Computer Science and Engineering, Seoul National University, Seoul 08826, South Korea}
\affiliation{Automation and Systems Research Institute, Seoul National University, Seoul 08826, South Korea}
\author{Beomgeun Cho}
\affiliation{Dept. of Computer Science and Engineering, Seoul National University, Seoul 08826, South Korea}
\affiliation{Automation and Systems Research Institute, Seoul National University, Seoul 08826, South Korea}
\author{KwangYeul Choi}
\affiliation{Dept. of Computer Science and Engineering, Seoul National University, Seoul 08826, South Korea}
\affiliation{Automation and Systems Research Institute, Seoul National University, Seoul 08826, South Korea}
\affiliation{Inter-University Semiconductor Research Center, Seoul National University, Seoul 08826, South Korea}
\author{SeungWoo Yoo}
\affiliation{Dept. of Computer Science and Engineering, Seoul National University, Seoul 08826, South Korea}
\affiliation{Automation and Systems Research Institute, Seoul National University, Seoul 08826, South Korea}
\affiliation{Inter-University Semiconductor Research Center, Seoul National University, Seoul 08826, South Korea}
\author{Changhyun Jung}
\affiliation{Automation and Systems Research Institute, Seoul National University, Seoul 08826, South Korea}
\affiliation{Inter-University Semiconductor Research Center, Seoul National University, Seoul 08826, South Korea}
\affiliation{Dept. of Electrical and Computer Engineering, Seoul National University, Seoul 08826, South Korea}
\author{Junho Jeong}
\affiliation{Automation and Systems Research Institute, Seoul National University, Seoul 08826, South Korea}
\affiliation{Inter-University Semiconductor Research Center, Seoul National University, Seoul 08826, South Korea}
\affiliation{Dept. of Electrical and Computer Engineering, Seoul National University, Seoul 08826, South Korea}
\author{Changsoon Kim}
\affiliation{Inter-University Semiconductor Research Center, Seoul National University, Seoul 08826, South Korea}
\affiliation{Dept. of Intelligence and Information, Seoul National University, Seoul 08826, South Korea}
\author{Dong-Il ``Dan'' Cho}
\affiliation{Automation and Systems Research Institute, Seoul National University, Seoul 08826, South Korea}
\affiliation{Inter-University Semiconductor Research Center, Seoul National University, Seoul 08826, South Korea}
\affiliation{Dept. of Electrical and Computer Engineering, Seoul National University, Seoul 08826, South Korea}
\author{Taehyun Kim}
\thanks{taehyun@snu.ac.kr}
\affiliation{Dept. of Computer Science and Engineering, Seoul National University, Seoul 08826, South Korea}
\affiliation{Automation and Systems Research Institute, Seoul National University, Seoul 08826, South Korea}
\affiliation{Institute of Computer Technology, Seoul National University, Seoul 08826, South Korea}
\affiliation{Inter-University Semiconductor Research Center, Seoul National University, Seoul 08826, South Korea}
\affiliation{Institute of Applied Physics, Seoul National University, Seoul 08826, South Korea}

\begin{abstract}
Ion trap systems built upon microfabricated chips have emerged as a promising platform for quantum computing to achieve reproducible and scalable structures. However, photo-induced charging of materials in such chips can generate undesired stray electric fields that disrupt the quantum state of the ion, limiting high-fidelity quantum control essential for practical quantum computing. While crude understanding of the phenomena has been gained heuristically over the past years, explanations for the microscopic mechanism of photo-generated charge carrier dynamics remains largely elusive. Here, we present a photo-induced charging model for semiconductors, whose verification is enabled by a systematic interaction between trapped ions and photo-induced stray fields from exposed silicon surfaces in our chip. We use motion-sensitive qubit transitions to directly characterize the stray field and analyze its effect on the quantum dynamics of the trapped ion. In contrast to incoherent errors arising from the thermal motion of the ion, coherent errors are induced by the stray field, whose effect is significantly imprinted during the quantum control of the ion. These errors are investigated in depth and methods to mitigate them are discussed. Finally, we extend the implications of our study to other photo-induced charging mechanisms prevalent in ion traps.
\end{abstract}

\maketitle

%\linenumbers

%%%%%%%%%%%%%%%%%%%%%%%%%%%%%%%%%%%%%%%%%%
\section{Introduction}
\label{sec_intro}

Ion trap systems are rapidly scaling up as platforms for universal quantum computing by incorporating semiconductor fabrication technologies \cite{stick_ion_2006, bermudez_assessing_2017, jain_scalable_2020, romaszko_2020, akhtar_high-fidelity_2023}. Compact, miniaturized chips enable greater ion densities, increased flexibility in ion configurations via ion transport \cite{pino_2021}, and serve as test-beds for on-chip integrated optics \cite{niffenegger_integrated_2020, mehta_integrated_2020, ivory_integrated_2021}.

However, the proximity of the ions with surrounding materials causes the ions to be significantly susceptible to stray electric fields. Stray fields can be primarily categorized based on their underlying sources. First, there is field noise originating from thermal fluctuation and dissipation of charges on the material surface, which primarily causes ion heating and motional dephasing \cite{turchette_heating_2000, turchette_decoherence_2000, schriefl_decoherence_2006, safavi-naini_microscopic_2011, kumph_electric-field_2016, boldin_measuring_2018}. On the other hand, stray fields may arise from the excitation and subsequent dynamics of unpaired/excess charge carriers in the material, for instance, through photo-generation by scattered light. These fields cause ion displacement, leading to phase-modulated interactions with lasers due to excess micromotion \cite{lee_micromotion_2023}, drifts in secular frequencies \cite{ivory_integrated_2021}, and fluctuations in the Rabi frequency attributed to motion within finite beam widths, which are all detrimental to the motion-sensitive quantum operations.

Describing photo-induced charging processes at the microscopic level is a non-trivial task. It requires understanding the optical excitation channel and subsequent carrier dynamics in the chip's constituent materials, which can be crudely categorized as conductors, insulators, and semiconductors. Among these, the most frequently observed form of charging, commonly termed as dielectric charging, typically occurs in insulators or insulator/conductor structures \cite{harlander_trapped-ion_2010, wang_laser-induced_2011, harter_long-term_2014, doret_controlling_2012, hong_new_2017, ivory_integrated_2021, jung_2023}. Boundaries defined by the surfaces and interfaces between materials, along with inhomogeneities within a single material, both complicate the dynamics and contribute significantly to the overall process \cite{kronik_surface_1999, harlander_trapped-ion_2010}. Moreover, since ions can experience stray fields from merely 10 -- 1000 elementary charges on the chip surface \cite{harlander_trapped-ion_2010, jung_2023}, numerous processes can occur simultaneously, making it challenging to identify the dominant mechanism.

\begin{figure*}[ht]
\centering
\includegraphics[width=0.8\textwidth]{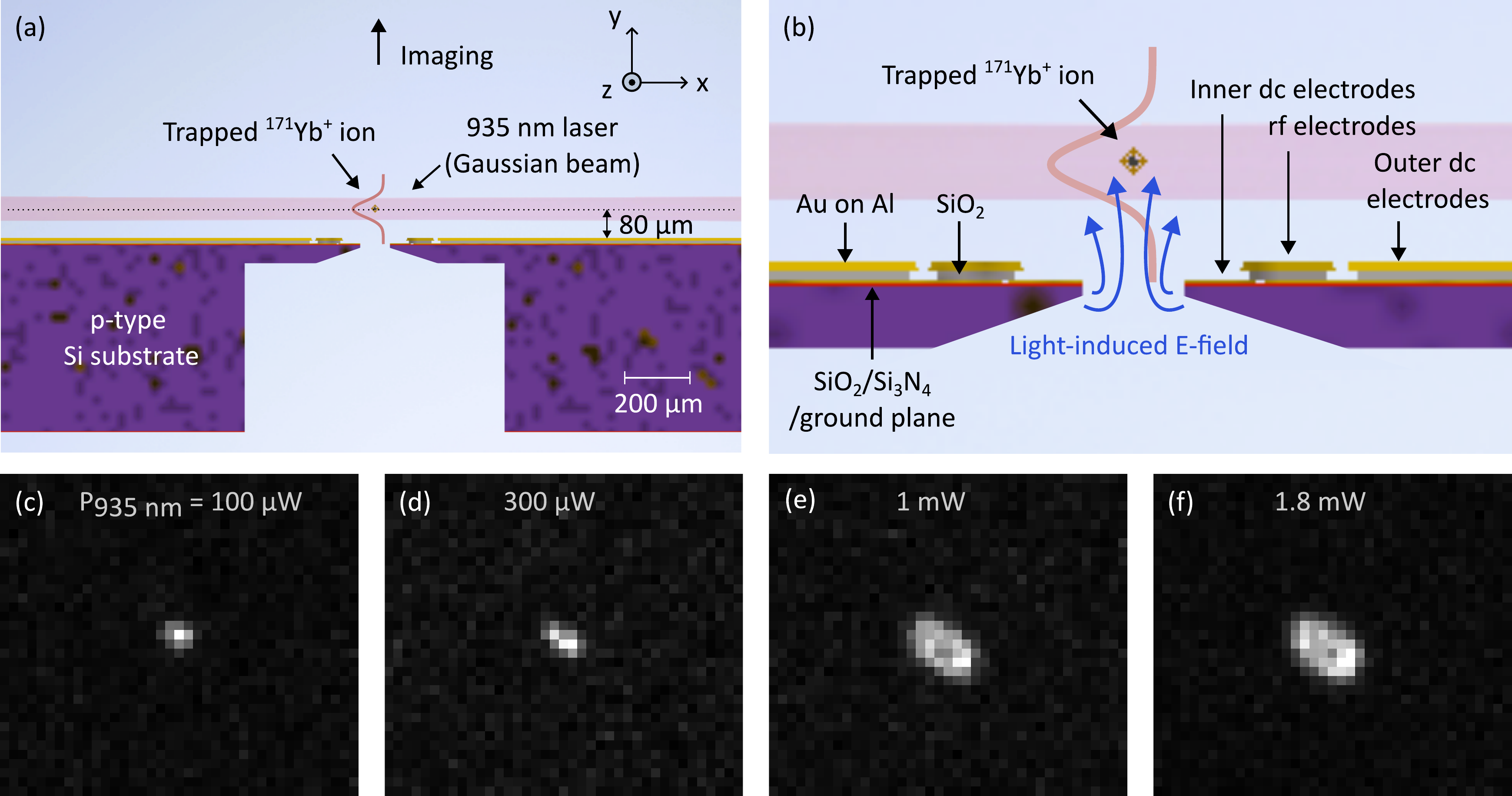}
\caption{Ion displacement by photo-induced electric field. (a) Simplified scheme of our chip trap structure near trap region with 935-nm laser. (b) An enlarged view of the trapping region with a schematic description of the photo-induced electric field. (c)-(f) Images of the ion displaced by the stray field from silicon substrate induced by scattered light as the power of the 935-nm laser was increased. (Color scale for each image was adjusted for better visibility.)}
\label{fig_ion_shift}
\end{figure*}

Due to such difficulties, the primary emphasis has been on mitigating these effects based on phenomenological observations rather than fully comprehending the carrier dynamics. Approaches include selecting materials with lower charging properties \cite{brown_materials_2021} or shielding materials prone to charging with metal coatings \cite{hong_new_2017,stick_demonstration_2010, jung_microfabricated_2021, blain_hybrid_2021}. To scale up ion trap chips with increasingly sophisticated features, however, a deeper understanding of these effects is essential as they require integration of heterogeneous materials into more intricate structures. Among the least understood phenomena is the carrier dynamics in semiconductors, particularly silicon \cite{niedermayr_cryogenic_2015, lakhmanskiy_heating_2019, blain_hybrid_2021}, which is increasingly favored as the substrate for these chips.

Our in-house fabricated micro-electromechanical systems (MEMS) chip and experimental setup serves as an excellent environment for directly investigating the stray fields generated from photo-induced charge carriers in silicon. A strong, systematic interaction between the ions and stray fields was first encountered with a near-infrared (NIR) laser at a wavelength of 935 nm. This observation alone sharply contrasts with numerous experimental reports on charging in ion traps, where the charging lasers typically lie in the ultraviolet (UV) or possibly visible (VIS) wavelengths \cite{harlander_trapped-ion_2010, wang_laser-induced_2011, harter_long-term_2014, doret_controlling_2012, hong_new_2017, mehta_integrated_2020, ivory_integrated_2021, jung_2023}, emphasizing the need for a novel model. 

To demonstrate the NIR charging of silicon, we inject a repumping 935-nm laser (with a maximum power of 1.8 mW and a waist of 45 \textmu m, traveling along $\hat{x}+\hat{z}$) as shown in Fig.~\ref{fig_ion_shift} (a). We then capture images of the ion while progressively ramping up the intensity of the laser, which consequently increases the scattered light. The images of the ion with different injected powers are shown from  Fig.~\ref{fig_ion_shift} (c)--(f), where defocus of the ion image caused by vertical displacement can be seen, especially between $P_{\textrm{935~nm}} =$ 300 \textmu W and 1 mW. We estimate the ion displacement by translating the electron-multiplying charge-coupled device (EMCCD) until the ion is in focus again. For a laser power of 1.8 mW, we measure an ion displacement of 8 \textmu m. The ion undergoes rapid displacement synchronized with the switching of the incident laser and immediately returns to its equilibrium position in the absence of light, with identical characteristic time scales for both processes. A comparable observation with blue lasers has been reported, but its underlying mechanism is unexplained \cite{mehta_ion_2014}.

In this study, we introduce a photo-induced charging model for semiconductors and utilize it to assess the surface conditions of the silicon substrate in our chip by using a $^{171}$Yb$^+$ ion as a quantum sensor. The spectral characteristics, photon flux dependence, and temporal evolution of the stray fields are measured and analyzed with motion-sensitive qubit transitions, and are accurately replicated through numerical simulations based on the proposed model. Then, the effect of the stray fields on the quantum dynamics of the trapped ion is analyzed through theoretical calculations using the Lindblad master equation. Our analysis confirms that stray fields induce coherent errors in the evolution of the ion, which is distinct from incoherent errors associated with the ion's thermal motion. Quantum control sequences to mitigate this error is presented. Finally, we extend the insights of our model to the more commonly reported dielectric charging phenomena and some studies on silicon charging in ion traps \cite{niedermayr_cryogenic_2014, lakhmanskiy_heating_2019, mehta_integrated_2020}, along with implications of the model in the context of fabrication.

%%%%%%%%%%%%%%%%%%%%%%%%%%%%%%%%%%%%%%%%%%
\section{Semiconductor charging model} 
\label{sec_model}

\begin{figure*}[ht]
\centering
\includegraphics[width=0.75\textwidth]{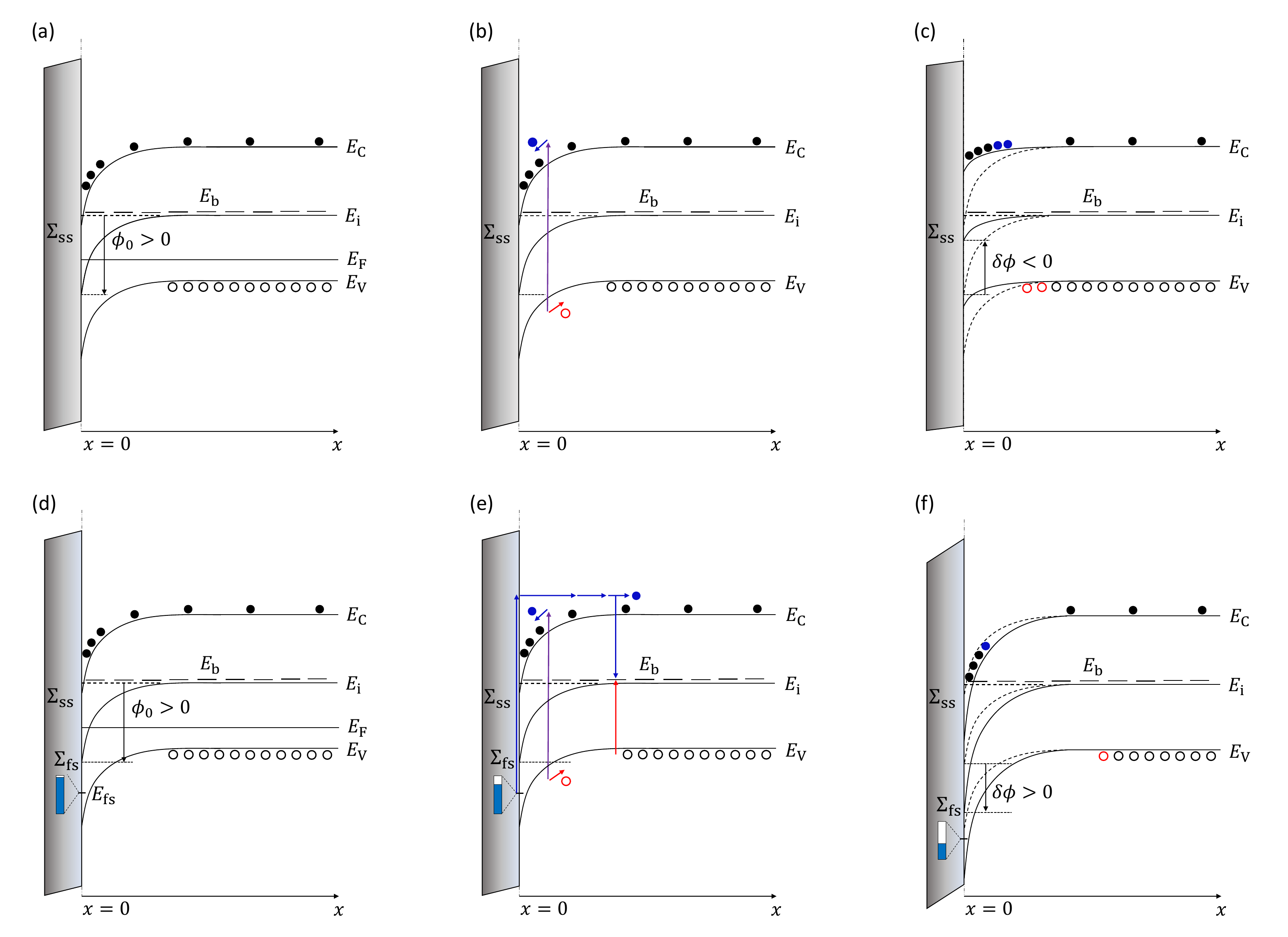}
\caption{Band diagram representation of the photoconductive charging model. The semiconductor has surfaces at $x=0$ and $x=l$, while the layer to the left of $x=0$ represents a native oxide layer. (a)--(c) Charging in the presence of fixed surface charges only (super-bandgap SPV). (a) The system in thermal equilibrium. The filled (empty) circles represent electrons (holes). (b) Carrier dynamics in (thermal) non-equilibrium. Photo-generated electrons (blue) are attracted to the surface and holes (red) are repelled into the bulk. (c) The steady state of the system in non-equilibrium. A negative SPV is formed as a result of the charge distribution. (d)--(f) Charging in the presence of both fixed surface charges and interface states (SPV inversion). (d) The system in thermal equilibrium. The blue color bar represents the electron occupation probability of the interface state. (e) Carrier dynamics in (thermal) non-equilibrium. In addition to the charge distribution process explained in (b), electrons optically excited from the interface state can diffuse into the bulk and recombine with free holes via bulk defect states. (f) The steady state of the system in non-equilibrium. A positive SPV is developed as a result of the depletion of holes at the surface. $E_{\mathrm{C}}$ and $ E_{\mathrm{V}}$ indicate the conduction and valence band edges, $E_{\mathrm{i}}$ and $\ E_{\mathrm{F}}$ the mid-bandgap and Fermi level, and $E_{\mathrm{b}}$ and $ E_{\mathrm{fs}}$ the energy levels of the bulk defect state and the interface state.}
\label{fig_model}
\end{figure*}

We present a model for the generation and distribution of carriers in a semiconductor under illumination, which describes the stray field experienced by the ion. The model is established within the framework of surface photovoltage (SPV) theory \cite{kronik_surface_1999, schroder_surface_2001, garrett_physical_1955, johnson_large-signal_1958, lagowski_photovoltage_1971}, utilizing the semiconductor equations to analyze the modification of surface potentials induced by light \cite{roosbroeck_sem_1950} (see Appendix~\ref{appendix_sem_eq}). Our model does not assume local charge neutrality, which allows us to compute the steady state distribution of carriers that is responsible for the SPV.

Let us first consider the most common semiconductor charging mechanism where electrons and holes are photo-generated in the bulk through band-to-band transitions, and separated by an electric field set up by fixed surface charges (also addressed as slow surface states \cite{johnson_large-signal_1958}) located at the exterior of the semiconductor body (see Fig.~\ref{fig_model} (a)--(c)). Linear bulk photo-generation occurs with the rate $G_{\mathrm{b}}(x) = N_0\alpha_{\mathrm{b}}$ exp$(-\alpha_{\mathrm{b}} x)$, where $N_0$ is the incident photon flux, $\alpha_{\mathrm{b}}$ is the absorption coefficient, and the illuminated surface is located at $x=0$. Silicon surfaces can easily become oxidized, acting as sites for hosting fixed oxide charges with a typical surface density of $\Sigma_{\mathrm{ss}} = +1 \times 10^{11}$ cm$^{-2}$ \cite{nicollian_sio2_1967, cheng_ox_1977}. In order to reflect such realistic surface conditions into our model, a native oxide layer is presumed to have formed on the exposed surfaces of our substrate shown in Fig.~\ref{fig_ion_shift} (b) by exposure to the atmosphere. In particular, this produces a surface potential of $\phi_0\approx$ +0.64 V \cite{kingston_calculation_1955} for a p-type doping concentration of $10^{15}$ cm$^{-3}$ (Fig.~\ref{fig_model} (a)). When the substrate is illuminated, photo-generated electrons are attracted to the surface and holes are repelled into the bulk due to drift under the surface electric field (Fig.~\ref{fig_model} (b)). Such distribution of charges screens the field (reduced downward band bending), leading to a change in SPV $\delta \phi$ that diminishes the initial surface potential $\phi_0$ (Fig.~\ref{fig_model} (c)). For p-type silicon, this SPV is usually negative and eventually saturates at $\delta \phi=-\phi_0$ for a sufficiently high photon flux \cite{kronik_surface_1999, schroder_surface_2001}. This is typically referred to as super-bandgap SPV \cite{kronik_surface_1999}. However, the SPV observed in our system is positive and can surpass $\phi_0$ in magnitude. Therefore, the described mechanism is not compatible with the experimental results.

To better account for the observation, we introduce interface states (or fast surface states \cite{garrett_fast_1956, johnson_large-signal_1958}) originating from surface defects localized near the silicon and oxide interface, which can act as centers for photo-induced defect-to-band transitions and surface recombination \cite{bebb_ocs_1967, chaudhuri_cross_1982, hsieh_recomb_1989, leibovitch_surf_1994} (see Fig.~\ref{fig_model} (d)--(f)). Unlike typical interface states formed at the Si-SiO$_2$ interface \cite{cheng_ox_1977}, these defect states are presumed to have formed by a deep reactive ion etching (RIE) process through which the silicon substrate was etched by bombardment of highly energetic plasma composed of various chemical gases such as carbon fluoride, sulfur fluoride, and argon \cite{jung_microfabricated_2021}. Such mechanically modified surfaces are known to host numerous surface states \cite{garrett_fast_1956}. Moreover, many types of surface defects have been reported in deep-level transient spectroscopy studies with samples etched by RIE processes \cite{matsumoto_rie_1982, gatzert_rie_2006}, and some have been directly related to SPV effects \cite{mohammed_surf_2005}. The interface state in our system is assumed to be donor-type (defined in Appendix~\ref{appendix_surf}) with a surface density $\Sigma_{\mathrm{fs}}$ and an effective single energy level $E_{\mathrm{fs}}$ below the Fermi level $E_{\mathrm{F}}$, as shown in Fig.~\ref{fig_model} (d). The interface exhibits a charge density of +$e(1-f_{\mathrm{s}}) \Sigma_{\mathrm{fs}}$, where $f_{\mathrm{s}}$ is the electron occupation probability of the interface states.

Under illumination, electrons in the interface states are optically excited to the conduction band to form a highly localized concentration near the interface. While some electrons are captured back into the defects (surface recombination), others diffuse into the bulk due to a large gradient in the density. In the bulk, they recombine with free holes via bulk defect states through the Shockley-Read-Hall recombination process, extending the surface depletion range (see Fig.~\ref{fig_model} (e)). The depletion of holes in the bulk is balanced by positive charging of the surface (enhanced downward band bending). Given a sufficiently high $\Sigma_{\mathrm{fs}}$, the positive charging enhances with stronger absorption of light and lower recombination of carriers occurring at these defects. A positively charged steady state is then established by the balance between the diffusion of electrons from the surface into the bulk, and the screening behavior of photo-generated bulk electrons. This can result in a large and positive SPV (denoted as $\delta \phi > 0$ in Fig.~\ref{fig_model} (f)), and is referred to as SPV inversion \cite{lagowski_photovoltage_1971, maltby_inversion_1975, kronik_surface_1999}. 

The steady state value of the occupation probability of the interface state $\bar{f_\mathrm{s}}$ is determined by the electron and hole densities $n$ and $p$ at the surface, and parameters that characterize optical absorption and surface recombination (see Appendix~\ref{appendix_surf}) \cite{shockley_recomb_1952, hall_recomb_1952, hsieh_recomb_1989}:
\begin{equation}
\bar{f_s}=\left. \frac{n+(s_{\mathrm{p0}}/s_{\mathrm{n0}})p_{\mathrm{1,s}}}{n+n_{\mathrm{1,s}} + (s_{\mathrm{p0}}/s_{\mathrm{n0}})\left(p+p_{\mathrm{1,s}}\right) + N_0\alpha_\mathrm{n}/s_{\mathrm{n0}}}\right|_{x=0}
\label{eqn_fs}
\end{equation}
where $s_{\mathrm{n0}}=\sigma_{\mathrm{n,s}}^{\mathrm{c}} \Sigma_{\mathrm{fs}} v_{\mathrm{n}}$ and $ s_{\mathrm{p0}}=\sigma_{\mathrm{p,s}}^{\mathrm{c}} \Sigma_{\mathrm{fs}} v_{\mathrm{p}}$ are the surface recombination velocities with the carrier capture cross sections $\sigma_{\mathrm{n,s}}^{\mathrm{c}}$, $\sigma_{\mathrm{p,s}}^{\mathrm{c}}$, and the bulk thermal velocities $v_{\mathrm{n}}$, $v_{\mathrm{p}}$. Also, $n_{\mathrm{1,s}}=n_{\mathrm{i}}$exp$(E_{\mathrm{fs}}/kT),\ p_{\mathrm{1,s}}=n_{\mathrm{i}}$exp$(-E_{\mathrm{fs}}/kT)$ where $n_{\mathrm{i}}$ is the intrinsic carrier concentration, $E_{\mathrm{fs}}$  is the energy level of the interface state, $k$ is the Boltzmann constant, and $T$ is the temperature of the system. The most significant parameter is $\alpha_\mathrm{n}$, the surface absorption coefficient, defined as $\alpha_\mathrm{n}=\sigma_\mathrm{n}^\mathrm{o} \Sigma_{\mathrm{fs}}$. It is proportional to the optical cross section, $\sigma_\mathrm{n}^\mathrm{o}$, whose value and spectral dependence can result in very different SPV effects as compared to when only bulk absorption is present. 

There are many theories on the microscopic origin of the optical cross section of impurities in semiconductors \cite{lucovsky_photoionization_1965, bebb_ocs_1967, anderson_ocs_1975, stoneham_phonon_1979, olenjnikova_ocs_1981, chaudhuri_cross_1982, coon_green_1986, ilaiwi_ocs_1990, tomak_ocs_1982}. The common objective is to find a suitable potential for the bound state $\ket{i}$ that best reproduces the observed spectral response through the dipole transition (i.e., electric field polarized in the z direction)
\begin{equation}
\sigma_\mathrm{n}^\mathrm{o} \propto
\left|\sum\nolimits_{f} \bra{f} \hat{z} \ket{i}\right|^{2} \delta(E-E_\mathrm{io})
\end{equation}
where $\ket{f}$'s are the continuum states in the conduction bands, and $E_{\mathrm{io}} = E_{f} - E_{i}$ is the ionization energy between the relevant conduction band and the defect level. Here, we apply the Hulth\'{e}n potential \cite{ilaiwi_ocs_1990}
\begin{equation}
\label{eqn_hulthen}
V(x) = -\frac{e^2}{\epsilon} \frac{\lambda}{a} \frac{e^{-\lambda x/a}}{1 - e^{-\lambda x/a}}
\end{equation}
which is basically a screened Coulomb potential, appropriate for describing shallow bound states, $\ket{i}$. The reason for choosing this potential will become evident in Sec.~\ref{sec_null_shift} where fitted values for the parameters $E_\mathrm{io},\ a$, and $\lambda$ are presented along with experimental data. $\epsilon$ is the dielectric constant of the material. Note that Eq.~(\ref{eqn_fs}) describes the optical excitation of electrons only. For the complete formalism also involving excitation of holes into the interface state (band-to-defect transition), see Appendix~\ref{appendix_surf}.

In the bulk, the dominant bulk recombination process in indirect semiconductors such as silicon is the Shockley-Read-Hall type, described by the rate
\begin{equation}
R_{\mathrm{b}}=\frac{np-n_\mathrm{i}^{2}}{\tau_{\mathrm{n0}}(p+p_{\mathrm{1,b}})+\tau_{\mathrm{p0}}(n+n_{\mathrm{1,b}})}
\end{equation}
where $n_{\mathrm{1,b}}=n_{\mathrm{i}}$ exp$(E_{\mathrm{b}}/kT)$, $p_{\mathrm{1,b}}=n_{\mathrm{i}}$ exp$(-E_{\mathrm{b}}/kT)$, with $E_\mathrm{b}$ denoting the energy level of the bulk defect. The lifetime parameters  $\tau_{\mathrm{n0}}=(\sigma_{\mathrm{n,b}}^c N_{\mathrm{b}} v_\mathrm{n} )^{-1}$ and $\tau_{\mathrm{p0}}=(\sigma_{\mathrm{p,b}}^\mathrm{c} N_\mathrm{b} v_\mathrm{p} )^{-1}$ characterize properties of the bulk defect recombination center. Here, $\sigma_{\mathrm{n,b}}^\mathrm{c}$ and $ \sigma_{\mathrm{p,b}}^\mathrm{c}$  are the capture cross sections of the bulk defect state for electrons and holes, respectively, and $N_\mathrm{b}$ is its density. The contribution of $R_\mathrm{b}$ is significant to the observed SPV effect, as it determines the degree to which holes are depleted.

\begin{figure*}[ht]
\centering
\includegraphics[width=1\textwidth]{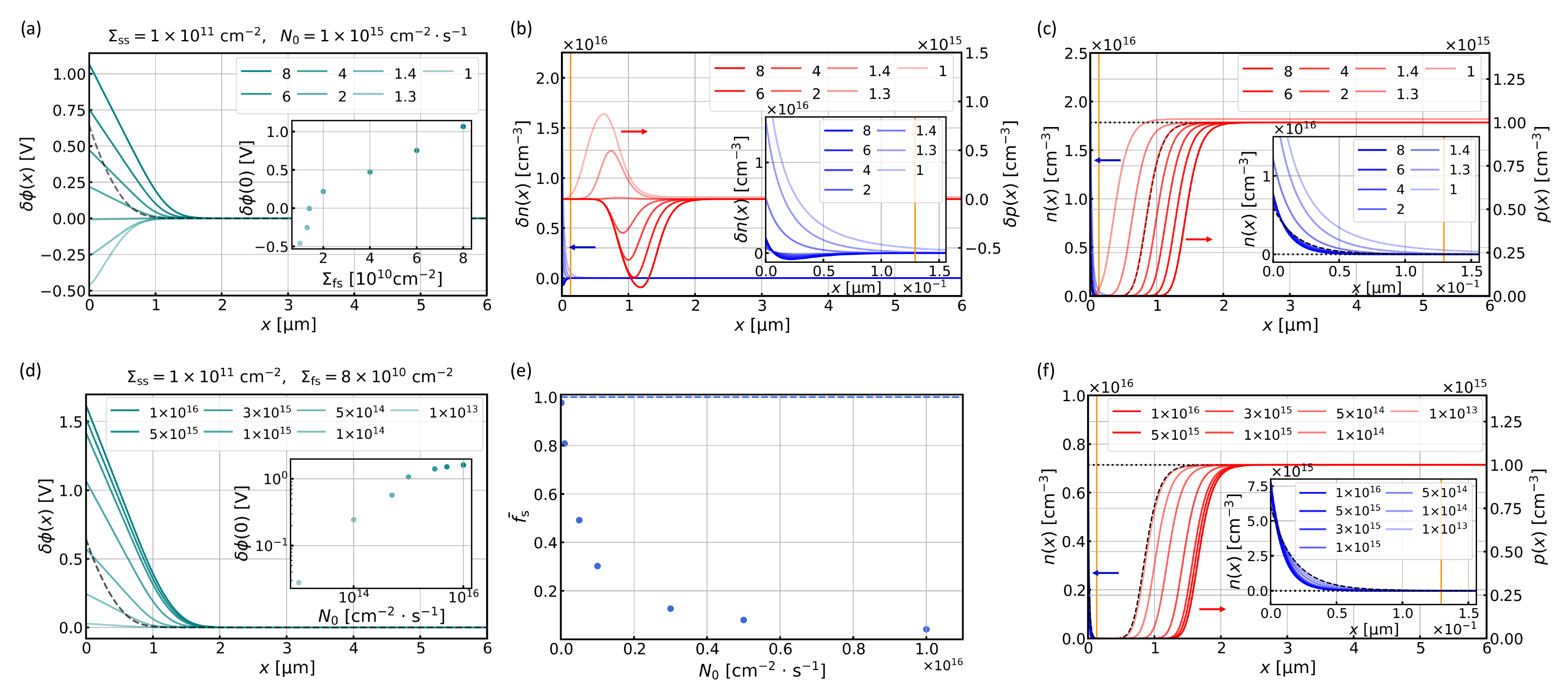}
\caption{Numerical simulation results of the photoconductive charging model. (a)--(c) Spatial profiles of (a) $\delta \phi$ (dark cyan), $\phi_0$ (black dashed), (b) $\delta n$ (blue), $\delta p$ (red), (c) $n=n_0+\delta n$ (blue), $p=p_0+\delta p$ (red), and $n_0,\ p_0$ (black dashed), for various $\Sigma_{\mathrm{fs}}$ with fixed photon flux $N_0=1\times10^{15}$ cm$^{-2} \cdot$ s$^{-1}$. Values of $\Sigma_{\mathrm{fs}}$ are denoted in the legends in units of 10$^{10}$ cm$^{-2}$. Inset in (a) shows values of $\delta \phi$ at $x=0$ for different $\Sigma_{\mathrm{fs}}$'s. From (b)--(c), it can be seen that the distinctive characteristic of a positive (negative) SPV is the widening (narrowing) of the surface depletion range in comparison to its range in thermal equilibrium. The orange vertical lines at $x=0.129$ \textmu m correspond to the Debye-screening length (defined as $r_+$ in Appendix~\ref{appendix_bulk_1}), which indicates that excess electrons in the bulk are mostly involved in screening. (d)--(e) Spatial profiles of (d) $\delta \phi$ (darkcyan), $\phi_0$ (black dashed), (e) $\bar{f_\mathrm{s}}$ (circle), $\bar{f_\mathrm{s}}_0 \approx 1$ (dashed), (f) $n$ (blue), and $p$ (red), for various photon fluxes $N_{0}$'s and fixed $\Sigma_{\mathrm{fs}}=8\times10^{10}$ cm$^{-2}$. Inset in (d) shows values of $\delta \phi$ at $x=0$ for different $N_0$'s. Saturation of SPV and surface depletion range appear as $\bar{f_\mathrm{s}}$ approaches 0. Dotted horizontal lines in (c) and (f) indicate $n_0=10^5$ cm$^{-3}$ (inset) and $p_0=10^{15}$ cm$^{-3}$, for the case when surface charges and interface states are absent (flat initial bands). Throughout (a)--(f), $\Sigma_{\mathrm{ss}}=1\times10^{11}$ cm$^{-2}$, while the remaining parameter values were chosen and fixed to best illustrate the described effects.
}
\label{fig_model_sim}
\end{figure*}

Numerical simulations to verify the proposed charging mechanism were performed by solving the semiconductor equations (see Appendix~\ref{appendix_sem_eq}). Bulk conditions were set in accordance with our chip substrate, which is p-type silicon doped at a concentration of $10^{15}$ cm$^{-3}$. The results are shown in Fig.~\ref{fig_model_sim}. A positive SPV occurs as predicted, increasing in magnitude with larger depopulation of electrons from the interface state, $\delta f_\mathrm{s} = \bar{f_\mathrm{s}} - \bar{f_\mathrm{s}}_0 < 0$, where $\bar{f_\mathrm{s}}_0$ is the electron occupation probability in thermal equilibrium. The carrier densities are decomposed as $n=n_0 + \delta n,\ p=p_0 + \delta p$ where $n_0, \ p_0$ are thermal equilibrium densities and $\delta n, \ \delta p$ are the excess densities. Fig.~\ref{fig_model_sim} (a)--(c) display the effect of an increasing interface state density $\Sigma_{\mathrm{fs}}$ under a fixed surface charge density $\Sigma_{\mathrm{ss}}=1\times10^{11}$ cm$^{-2}$ and constant photon flux. A noticeable inversion in the SPV takes place as $\Sigma_{\mathrm{fs}}$ increases, when charging from surface absorption outweighs that from bulk absorption. Fig.~\ref{fig_model_sim} (d)--(f) show the effect of an increasing photon flux while the charge densities were kept constant at $\Sigma_{\mathrm{fs}}=8\times10^{10}$ cm$^{-2}$  and $\Sigma_{\mathrm{ss}}=1\times10^{11}$ cm$^{-2}$. The correlation between the enhancement in SPV, depopulation of electrons $\delta \bar{f_\mathrm{s}}<0$, and the increasing depth of surface depletion verifies our charging model. 

In the following section, we briefly explain how the experimental value of the SPV can be determined using the ion as a quantum sensor for detecting stray fields arising from the SPV, and then display theoretical results, along with experimental data. 

\section{Measurement of SPV and its dependence on optical power and wavelength}
\label{sec_null_shift}

\begin{figure*}[ht]
\centering
\includegraphics[width=0.9\textwidth]{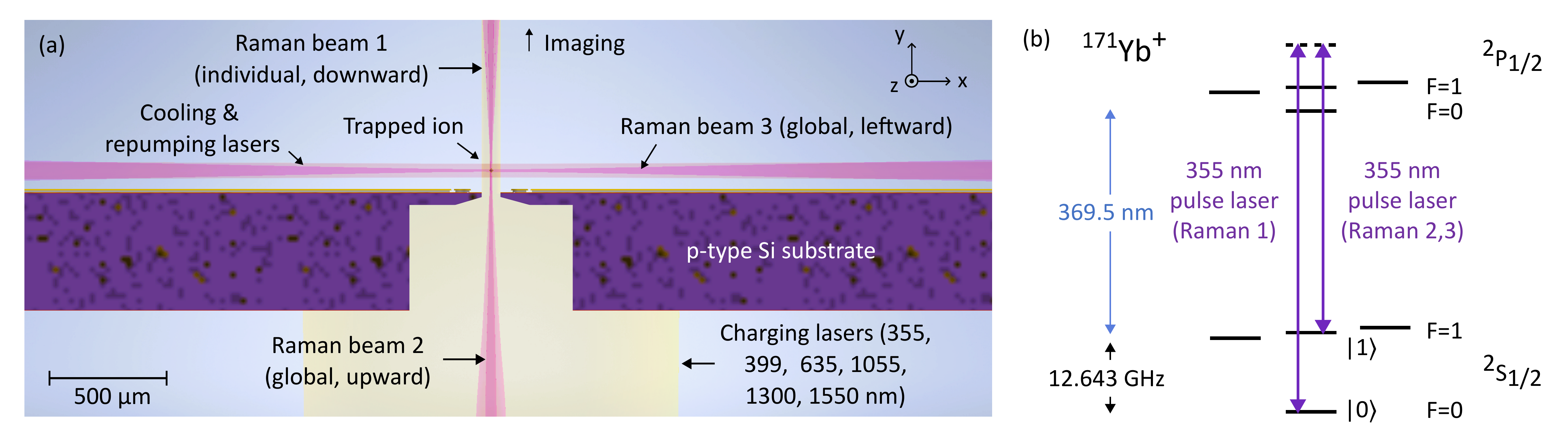}
\caption{Experimental configuration. (a) Schematic cross-sectional illustration of the microfabricated chip with parallel and perpendicularly incident laser beams. (b) Energy level diagram of $^{171}$Yb$^+$ with Raman transition for qubit control by pulsed laser beams.}
\label{fig_chip}
\end{figure*}

\begin{figure*}[t]
\centering
\includegraphics[width=0.75\textwidth]{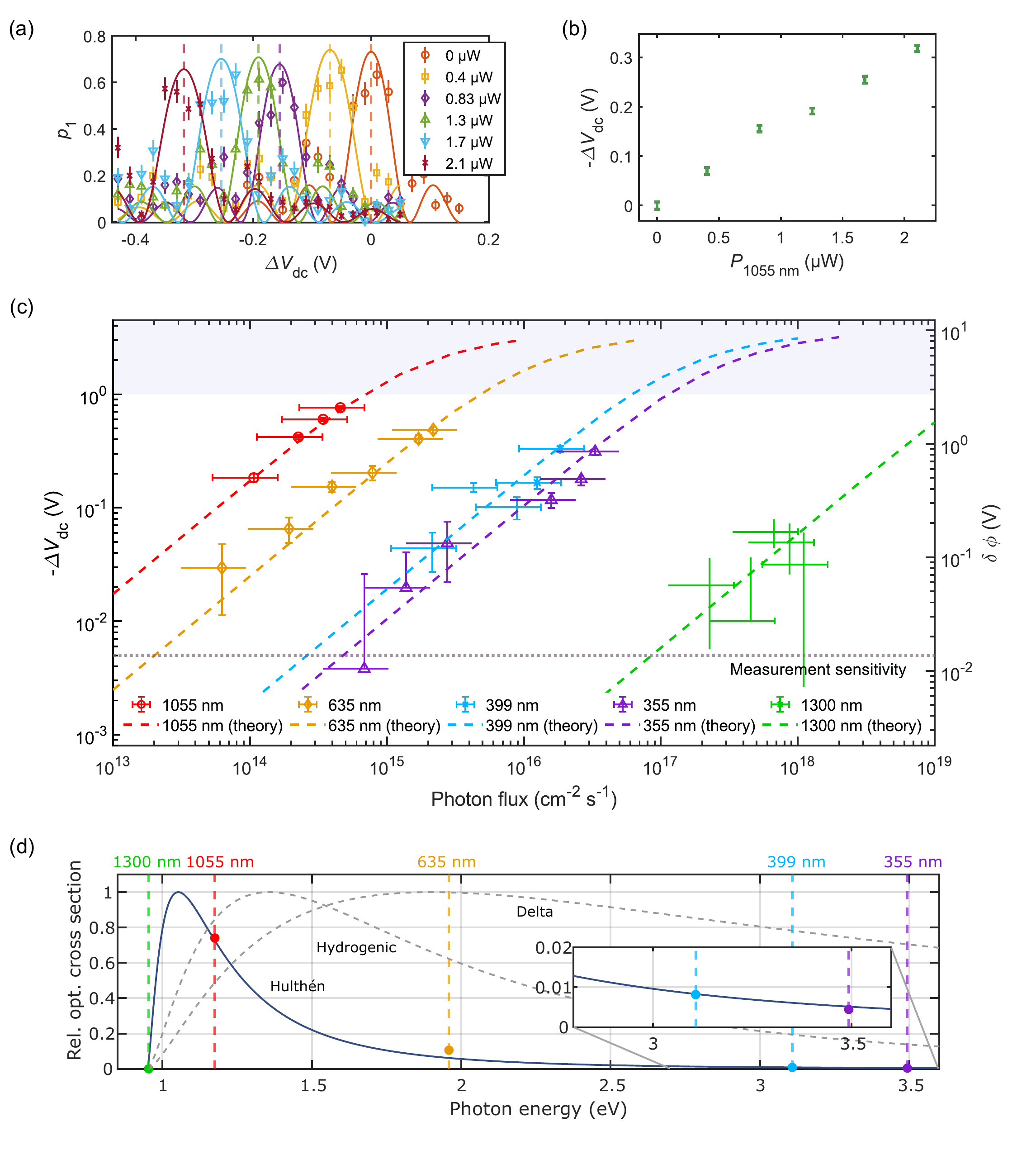}
\caption{Measurement for laser power and wavelength dependence of the laser-induced field magnitude. (a) Raman transition probabilities vs. inner dc voltage shift for a range of 1055-nm laser power. (b) Inner dc compensation voltage vs. 1055-nm laser power. (c) Spectral response of laser-induced electric fields with respect to the compensation voltage (left) and the SPV (right). The shaded region indicates the unstable trapping condition of the ion. (d) Normalized optical cross section vs. photon energy. The solid circles are the values used to fit the experimental data in (c), with each color corresponding to the respective wavelength. The solid line corresponds to the fitted curve for the Hulth\'{e}n potential in Eq.~(\ref{eqn_hulthen}) while the dashed lines indicate the two limiting cases from the quantum defect model \cite{anderson_ocs_1975, chaudhuri_cross_1982}. The vertical error bars in (a)--(c) indicate the 95 \% confidence intervals of the fit and the horizontal error bars indicate $\pm$50 \% of the photon flux values which reflects their overall uncertainties.}
\label{fig_power_wavelength}
\end{figure*}

\begin{table*}[ht]
\caption{Parameter values used in the numerical simulation}
\label{tbl_param}
\setlength{\tabcolsep}{2pt}
\centering
\begin{tabular}{cccccccccc}
\hline
\multirow{2}{*}{Surface} & \begin{tabular}[c]{@{}c@{}}$\Sigma_{\mathrm{fs}}$\\ {[}cm$^{-2}${]}\end{tabular} & \begin{tabular}[c]{@{}c@{}}$\sigma_\mathrm{n,s}^\mathrm{c}=\sigma_\mathrm{p,s}^\mathrm{c}=\sigma^\mathrm{c}_\mathrm{s}$\\ {[}cm$^{2}${]}\end{tabular}    & \begin{tabular}[c]{@{}c@{}}$E_{\mathrm{fs}}$\\ {[}eV{]}\end{tabular} & \begin{tabular}[c]{@{}c@{}}$\Sigma_{\mathrm{ss}}$\\ {[}cm$^{-2}${]}\end{tabular}                                                & \begin{tabular}[c]{@{}c@{}}$\sigma_\mathrm{n}^\mathrm{o}(1300)$\\ {[}cm$^{-2}${]}\end{tabular} & \begin{tabular}[c]{@{}c@{}}$\sigma_\mathrm{n}^\mathrm{o}(1055)$\\ {[}cm$^{2}${]}\end{tabular} & \begin{tabular}[c]{@{}c@{}}$\sigma_\mathrm{n}^\mathrm{o}(635)$\\ {[}cm$^{2}${]}\end{tabular} & \begin{tabular}[c]{@{}c@{}}$\sigma_\mathrm{n}^\mathrm{o}(399)$\\ {[}cm$^{2}${]}\end{tabular} & \begin{tabular}[c]{@{}c@{}}$\sigma_\mathrm{n}^\mathrm{o}(355)$\\ {[}cm$^{2}${]}\end{tabular} \\ \cline{2-10}
        &2.7$\times 10^{11}$                                              & 6.48$\times$$10^{-24}$                                                     & $-$ 0.39    & 1.0$\times 10^{11}$                                                                                             & 1.1$\times 10^{-19}$                                                   & 3.24$\times 10^{-15}$                                                   & 3.6$\times 10^{-16}$    & 3.6$\times 10^{-17}$ & 1.95$\times 10^{-17}$                                               \\ \hline
\multirow{2}{*}{Bulk}  & \begin{tabular}[c]{@{}c@{}}$N_\mathrm{b}$\\ {[}cm$^{-3}${]}\end{tabular}  & \begin{tabular}[c]{@{}c@{}}$\sigma_{\mathrm{n,b}}^c$\\ {[}cm$^{2}${]}\end{tabular} & \begin{tabular}[c]{@{}c@{}}$\sigma_{\mathrm{p,b}}^c$\\ {[}cm$^{2}${]}\end{tabular} & \begin{tabular}[c]{@{}c@{}}$E_{\mathrm{b}}$\\ {[}eV{]}\end{tabular} & \begin{tabular}[c]{@{}c@{}}$n_i$\\ {[}cm$^{-3}${]}\end{tabular}     & \multicolumn{2}{c}{\begin{tabular}[c]{@{}c@{}}$v_\mathrm{n}=v_\mathrm{p}=v_\mathrm{th}$\\ {[}cm$\cdot$ s$^{-1}${]}\end{tabular}} & \begin{tabular}[c]{@{}c@{}}$\mu_\mathrm{n}$\\ {[}cm$^{2} \cdot$ V$^{-1}\cdot$ s$^{-1}${]}\end{tabular}     & 
\begin{tabular}[c]{@{}c@{}}$\mu_\mathrm{p}$\\ {[}cm$^{2} \cdot$ V$^{-1}\cdot$ s$^{-1}${]}\end{tabular}    \\ \cline{2-10}
        &1.0$\times 10^{13}$                                               &1.0$\times 10^{-15}$                                                     &1.0$\times 10^{-15}$                                                     & 0.0    &1.0$\times$10$^{10}$     & \multicolumn{2}{c}{1.0$\times10^{7}$}                                                                                              & 1340                                                      & 284                                                     \\ \hline
\end{tabular}
\end{table*}

The magnitude of the SPV at the exposed surface of the silicon substrate is estimated using a micromotion compensation scheme. This scheme involves dc-scanning \cite{lee_micromotion_2023}, through which the stray field can be directly measured in terms of a compensation voltage. We can either measure the qubit transition rate with Raman beams 1 and 2 in Fig. \ref{fig_chip} (a) where the relevant energy levels are shown in Fig. \ref{fig_chip} (b), or the absorption rate of a weak 935-nm laser injected vertically to the chip. Both methods utilize motion-sensitive responses of the ion that display a Bessel-like profile where the maximum occurs at the compensation voltage that cancels out the stray field in the scanning direction. The voltages of the inner dc electrode pair, shown in Fig. \ref{fig_ion_shift} (b) are scanned to search for the optimal compensation field (refer to Ref. \cite{jung_microfabricated_2021} for more details of the chip). Since our experimental configuration enables the generation and compensation of stray fields only in the direction normal to the chip surface, the theoretical model is greatly simplified.

First, the dependence of the SPV on the power of illumination is presented. A 1055-nm laser beam is shone to uniformly illuminate the backside of the chip, as shown in Fig. \ref{fig_chip} (a). The qubit-flipping probabilities $p_1$ are measured with the dc voltages scanned under various powers of the laser from 0 -- 2.1 \textmu W, as shown in Fig.~\ref{fig_power_wavelength} (a). The extracted compensation voltages for various laser powers $P_{\textrm{1055~nm}}$ are plotted in Fig.~\ref{fig_power_wavelength} (b). The absolute value of compensation voltage grows linearly with increasing laser power. To estimate the internally generated SPV from the measured compensation voltage, the COMSOL Multiphysics\textsuperscript{\tiny\textregistered} software is employed for electrostatic analysis which shows that a compensation voltage of $\Delta V_{\mathrm{dc}} \approx -0.1$ V is required to cancel the stray field generated by an SPV of $\phi \approx +0.273$ V at the silicon surface (see Appendix \ref{appendix_sim}). The strength of the corresponding field is 288 V$\cdot$m$^{-1}$ at the position of the ion, which can displace the ion by 1.6 \textmu m for a harmonic oscillator with a secular frequency of 1.6 MHz. This implies that an SPV of merely several mV is strong enough to deteriorate the fidelity of quantum operations of trapped ions, where coherent displacements typically employed in M{\o}lmer--S{\o}rensen gates is a few times the ground state wave packet width which is around 4 nm \cite{campbell_ultrafast_2010}.

Next, the general spectral dependence of the SPV is obtained from the photo-induced response of the exposed silicon surface to six wavelengths: two in the UV range (355, 399 nm), one in the visible range (635 nm), and three in the NIR range (1055, 1300, and 1550 nm). The lasers are shone onto the backside of the chip with diameters of 1.9 -- 2.7 mm, fully illuminating the exposed silicon surface as in Fig.~\ref{fig_chip} (a). The compensation voltages against the photon fluxes of 355, 399, 635, 1055 and 1300-nm lasers are shown in Fig.~\ref{fig_power_wavelength} (c). Note that no measurable displacement of the ion was observed up to the maximum power of our 1550-nm laser (1.5 mW).

It can be clearly seen that the SPV is positive as indicated by the negative sign of the compensation voltage, which implies the occurrence of SPV inversion from a p-type silicon. Moreover, this behavior persists throughout all wavelengths, that is, not only at sub-bandgap but also super-bandgap wavelengths. To our knowledge, SPV inversion in silicon and SPV inversion at super-bandgap wavelengths have not been reported to date. The two phenomena, however, are simultaneously reproducible according to our semiconductor charging model under the hypothesized surface conditions from Sec.~\ref{sec_model}.

In particular, the unique dependence on the wavelength is largely determined by the optical cross section, as will be discussed in the following paragraph. The observation range of the ion displacement is bound from below by the measurement sensitivity of $|\Delta V_{\mathrm{dc}}| \approx$ 5 mV (from Ref. \cite{lee_micromotion_2023}, with consideration of probe laser intensity fluctuation; indicated by a dashed horizontal line in Fig.~\ref{fig_power_wavelength} (c)), and from above by the voltage that is required to stably confine the ion within the trap, which is $|\Delta V_{\mathrm{dc}}| \approx$ 1 V (shown as a shaded region in Fig.~\ref{fig_power_wavelength} (c)). The limited available powers of the diode lasers used in our experiments also set the upper bound for maximum displacement.

The theoretical curves of the SPV's fitted to experimental data are shown as dashed lines in Fig.~\ref{fig_power_wavelength} (c). The numerical values of the parameters used in the computation are listed in Table~\ref{tbl_param}. Typical values reported elsewhere \cite{siegfried_sem_2011, chenming_sem_2009, arora_mob_1982} were used for the bulk parameters, while the surface parameters were fitted to explain the experimental data. One exception is the fixed surface charge density, $\Sigma_{\mathrm{ss}}=1 \times 10^{11}$ cm$^{-2}$, which is the typical value observed in real oxidized silicon surfaces \cite{nicollian_sio2_1967}. The parameters $\mu_\mathrm{n}$ and $\mu_\mathrm{p}$ are the electron and hole mobilities that are necessary to solve the semiconductor equations (see Appendix~\ref{appendix_sem_eq}). The bulk absorption coefficient values used in simulation are, $\alpha_\mathrm{b}(1300)=2.7\times 10^{-5}$ cm$^{-1}$, $\alpha_\mathrm{b}(1055)=1.63\times 10^{1}$, cm$^{-1}$, $\alpha_\mathrm{b}(399)=9.52\times 10^{4}$ cm$^{-1}$ \cite{green_abs_1965}.

The optical cross sections are determined from the empirical spectral response of the SPV. The cutoff wavelength is found to be near 1300 nm ($E_\mathrm{io}\approx$ 0.95 eV). With $E_\mathrm{i}=0$ eV and the intrinsic bandgap of silicon being 1.12 eV, we have $E_\mathrm{fs}\approx-0.39$ eV, which lies lower than the Fermi level $E_\mathrm{F}\approx-0.3$ eV, determined by the doping concentration of our substrate. Lasers of wavelengths between 1055 -- 1300 nm will allow us to measure the cutoff more accurately, but the current estimation is sufficient for our purposes.

According to numerical simulations, the magnitude of the SPV at a particular wavelength is nearly proportional to the value of the optical cross section at that wavelength. The experimental values of the relative strengths between the optical responses at different wavelengths are plotted in Fig.~\ref{fig_power_wavelength} (d) as solid circles. The theoretical optical cross section, whose analytical expression can be found in Appendix~\ref{appendix_abs}, is fitted to these data points and overlaid on them. It is normalized relative to the peak value that occurs near the bandgap energy of silicon $\sim$ 1.12 eV. The fitted values for the Hulth\'{e}n potential parameters are $a=6.4\times10^{-8}$ cm, $a/\lambda=1\times10^{-7}$ cm. The absolute values of the optical cross section listed in Table~\ref{tbl_param} were obtained from this curve, and then used to numerically compute the theoretical SPV in Fig.~\ref{fig_power_wavelength} (c). Note that the optical cross section at 1300 nm is not exactly zero, but smaller than the peak value by several orders of magnitudes.

The effectiveness of the semiconductor charging model in explaining the observed spectral dependence of the SPV is mainly enabled by the Hulth\'{e}n potential, which is suitable for describing shallow defects. A characteristic feature of the optical response of shallow defects in semiconductors is a narrow absorption spectra near the bandgap energy \cite{boer_sem_2020}, which is indeed in agreement with experiment. The sharp contrast in the optical cross section between wavelengths in the NIR/VIS and UV ranges cannot be reproduced by the more common quantum defect models \cite{anderson_ocs_1975, chaudhuri_cross_1982}. The normalized optical cross section of two limiting cases in the quantum defect model, the delta-function and hydrogenic defects, are shown in Fig.~\ref{fig_power_wavelength} (d) for comparison. Also, for a fixed set of parameters, the optical cross section tends to become larger for shallower defects. In particular, the peak value of the optical cross section derived from the Hulth\'{e}n potential is larger than that predicted by the quantum defect models by 1 -- 2 orders in magnitude. 

The density and carrier capture cross sections of the interface state can then be determined as the set of values that best reproduce both the observed magnitude and slope of the SPV for a broad range of incident photon flux. Conditions for SPV inversion are found to favor identical or comparable values for the capture cross sections of electrons and holes, hence, $\sigma_\mathrm{n,s}^\mathrm{c}=\sigma_\mathrm{p,s}^\mathrm{c}=\sigma^\mathrm{c}_\mathrm{s}$. Under this condition, the last term in the denominator of Eq.~(\ref{eqn_fs}) becomes proportional to $\sigma_\mathrm{n}^\mathrm{o}/\sigma^\mathrm{c}_\mathrm{s} v_\mathrm{th}$. This ratio determines the sensitivity of the optical response, i.e., $\sigma_\mathrm{n}^\mathrm{o}(1055)/\sigma^\mathrm{c}_\mathrm{s} v_\mathrm{th}=50$ cm$^{-1} \cdot s$. Given the optical cross section values, the capture cross section is fitted to the very small value listed in Table~\ref{tbl_param}, which is also a property consistent with shallow defects \cite{alekperov_ccs_1998}. The large sensitivity is the primary reason for the peculiar SPV inversion observed in our system \cite{tong_1976}.

The experimental data can also be fitted well to the model in which an oxide layer is absent at the surface ($\Sigma_\mathrm{ss}=0$), with minor corrections to the parameter values. This condition is actually more favorable to the proposed surface charging process since bulk screening effects arising from carrier drift are reduced in the absence of positive fixed oxide charges. Therefore, the model's capability to effectively account for the two distinct surface conditions with relatively small adjustments to the parameter values demonstrates that the overall mechanism remains applicable to a broad range of uncertainties in the surface conditions. 

\section{Time-resolved Doppler shift}
\label{sec_frequency_shift}

\begin{figure}[h]
\centering
\includegraphics[width=1\columnwidth]{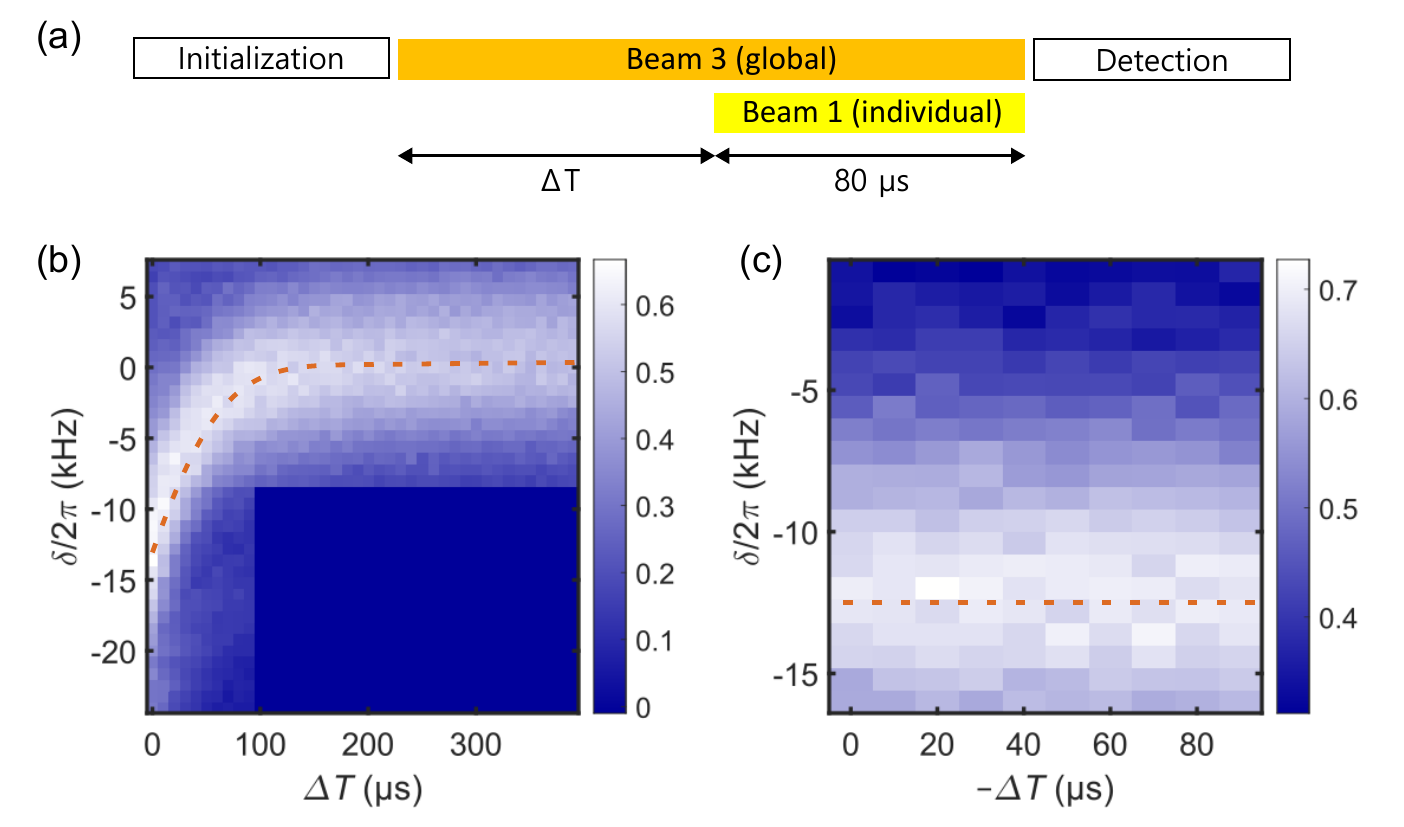}
\caption{Raman beam pre-turn-on measurement. (a) Operation sequence of pre-turn-on measurement for Raman beam 3. After state initialization, the global beam is first turned on, and then the individual beam is turned on after a time interval of $\Delta T$. Raman transition is driven for 80 \textmu s, followed by state detection. (b) Result of pre-turn-on measurement for Raman beam 3. Transition probability is plotted with varying $\Delta T$ and the laser detuning $\delta$. (c) Result of pre-turn-on measurement for Raman beam 1. The orange dashed lines serve as guides for the maximum values at each time interval.}
\label{fig_pre-turn-on}
\end{figure}

The characteristic time scale of the semiconductor charging is investigated by measuring the velocity of the ion driven by the photo-induced stray field originating from the scattering of the Raman beams used for quantum operations. In particular, the time-resolved Doppler shift of the resonant frequency of the Raman transition caused by Raman beam 3 (global) is measured. A schematic of the sequence is shown in Fig.~\ref{fig_pre-turn-on} (a). After state initialization, we turn on only Raman beam 3, which has a power of 360 mW (pre-turn-on as we call) and wait for varying time intervals $\Delta T$. Then we turn on Raman beam 1 whose power is 12 mW, to drive the Raman transition along with Raman beam 3, for a fixed time of 80 \textmu s which is close to the $\pi$-pulse duration. Finally, the qubit state of the ion is detected, and the average transition probability from repeated sequences is recorded. The Doppler shift between the ion and the Raman lasers is monitored by repetitions of the experiment for a range of detuning values $\delta$ of Raman beam 1 (by varying the rf frequency applied to the acousto-optic modulator).

The result of pre-turn-on measurement (for Raman beam 3) is shown in Fig.~\ref{fig_pre-turn-on} (b). It clearly reveals a continuous change of the resonance frequency of the transition in time and saturation after frequency shift of roughly 13 kHz. The gradual change in the qubit frequency implies that it is caused by the Doppler shift due to the movement of the ion. Assuming the following form of Doppler shift, $\Delta \omega_o = \Delta \vec{k} \cdot \vec{v}$, where $\Delta \vec{k}$ is the momentum difference of the two Raman beams and $\vec{v}$ is the ion velocity, the velocity is estimated to be 3.3 nm/\textmu s, which is significant considering that the width of the zero-point wavefunction is 4.3 nm for a secular frequency of 1.6 MHz of our trap and the wavelength of the Raman beatnote is 251 nm. In contrast, when Raman beam 1 is pre-turned on instead of Raman beam 3, no shift in the qubit frequency is observed as depicted in Fig.~\ref{fig_pre-turn-on} (c). The different results can be explained by the relatively low power of Raman beam 1 and the fact that it does not irradiate the exposed silicon directly. Note that the actual direction of the ion's velocity can be deviated from the chip's perpendicular direction, depending on the spatial profile of the SPV. Therefore, only the order of the speed was our consideration. 

Both the rise and fall time of the stray fields lie within the order of 1 -- 100 \textmu s, with slight variations depending on the laser alignment and power conditions. These time scales are comparable to that of the time evolution of the ion, capable of causing significant infidelity in operations involving motional quantum states. In uniform materials, the characteristic time scale of  neutralization is determined by the dielectric relaxation time $\tau_{+}$ which is estimated to be on the order of several pico-seconds for our substrate given the doping concentration of $10^{15}$ cm$^{-3}$ (see Appendix~\ref{appendix_bulk_2}). The measured time scales clearly do not agree with this value, but are instead on the order of carrier lifetimes associated with defects in semiconductors \cite{jiang_defect_2018}. This strongly supports our assertion that the observed SPV originates from charging processes associated with impurities throughout the surface and bulk of the semiconductor (see Appendix~\ref{appendix_bulk_2}). We will also use the dielectric relaxation time to interpret dielectric charging and other silicon charging phenomena in Sec.~\ref{sec_discussion}.

\section{Effects on quantum control}
\label{sec_treatment}

The motion of the ion in the presence of a rapidly developing stray field significantly modifies the motion-sensitive Rabi oscillation. According to our simulations, the main effect of the stray field is to induce coherent errors during quantum control. The error is coherent because the qubit state retains its purity but undergoes unintended unitary evolution, differing from the more common incoherent error originating from the thermal motion of the ion \cite{turchette_decoherence_2000}. We use the term thermal decoherence in a broad sense, involving dephasing arising from the thermal distribution of phonons, and heating from the environment.

The Rabi oscillation and trajectory of evolution on the Bloch sphere are simulated by solving the Lindblad master equation (see Appendix~\ref{appendix_lindblad}) whose system Hamiltonian is given as \cite{leibfried_quantum_2003}
\begin{equation}
\label{eqn_H_sys}
H_\mathrm{sys}(t) = \frac{\hbar \Omega}{2}\left(e^{i\Delta k x(t)}e^{-i\delta t}\sigma^{\dag} + e^{-i\Delta k x(t)}e^{i\delta t}\sigma \right) 
\end{equation}
where $\Omega$ is the Rabi frequency, $\sigma^{\dag} (\sigma)$ is the two level system raising (lowering) operator. The Lindblad operator for amplitude damping $L = \Gamma a $ is used where $\Gamma$ is the ion heating rate, $a^{\dag} (a)$ the raising(lowering) operator of the oscillator, with the temperature of the environment is assumed to be T=300 K. The one-dimensional Hamiltonian is consistent with the experimental configuration of the counter-propagating beam setup in Fig.~\ref{fig_chip}.

Assuming a stray field with an exponential temporal profile $E_\mathrm{str}(1-\mathrm{exp}(-t/\tau_\mathrm{str}))$ and a static compensation field $E_\mathrm{com}$, we can define the force exerted on the ion as
\begin{equation}
\label{eqn_field}
F(t) = e\left(E_\mathrm{str}\left(1-e^{-(t_\mathrm{pre}+t)/\tau_\mathrm{str}}\right) - E_\mathrm{com}\right)
\end{equation}
where $t_\mathrm{pre}$ is a pre-turn-on time that can be adjusted to control the initiation of the evolution. This stray field originates from the scattered light from Raman beam 2 (global) in the counter-propagating configuration. The effect of these fields have been absorbed into the position operator $x(t)$ as \cite{ji_1995, ji_1996} (see Appendix ~\ref{appendix_td_oscillator})
\begin{equation}
\label{eqn_pos}
x(t)\approx x_{0} \left(e^{i\omega_x t} a^{\dag} + e^{-i\omega_x t} a\right)+2x_{0} \mathrm{Re}(\alpha(t))
\end{equation}
where
\begin{multline}
\label{eqn_alpha}
\alpha(t) \approx e^{-i\omega_x t}\alpha(0) \\ 
+ \frac{i}{\hbar}\mathlarger{\mathlarger{\int}}_{0}^{t} dt' e^{-i\omega_x(t-t')}\left(\frac{1+\frac{q_x}{2}\mathrm{cos}(\omega_\mathrm{rf}t')}{1+\frac{q_x}{2}}\right)x_{0}F(t').
\end{multline}
Here, $x_{0}$ is the size of the ground state, $\omega_x, \ \omega_\mathrm{rf}$ indicate the secular and rf trap frequencies, respectively, and $q_x$ is the q-parameter associated with the trap stability \cite{leibfried_quantum_2003}. This expression results from directly solving the quantum equations of motion for the time-dependent oscillator in the presence of external fields \cite{ji_1996}. In particular, Eq.~(\ref{eqn_alpha}) represents the excess micromotion. The approximations in the above equations result from omitting a squeezing factor that describes intrinsic micromotion, which does not compromise the generality of the results.

The Rabi frequency and heating rate of our system are $\Omega=2\pi\times78$ kHz and $\Gamma\approx10^4$ quanta/s. The detuning is set to $\delta=0$, so that the carrier transition is driven. Also, $\omega_\mathrm{rf}=2\pi\times22.21$ MHz and $\omega_x=2\pi\times1.6$ MHz, from which we obtain $q_x \approx0.2$. The Lamb-Dicke factor is $\Delta k x_{0}\approx$ 0.152. In the Doppler limit, the mean phonon number $\bar{n}_0$ is calculated to be $\bar{n}_0\approx10$. This value may vary depending on the compensation of the stray field, due to modification of the cooling efficiency in the presence of excess micromotion \cite{berkeland_minimization_1998}, as is also observed in our experiments. Simulations are performed by treating $\tau_\mathrm{str},\ E_\mathrm{str},\ E_\mathrm{com}$, and $\bar{n}_0$ as fitting parameters, while the experimental parameters $\Omega,\ \Gamma,\ \omega_x$ and $q_x$ are fixed.

\begin{figure*}[t]
\centering
\includegraphics[width=1\textwidth]{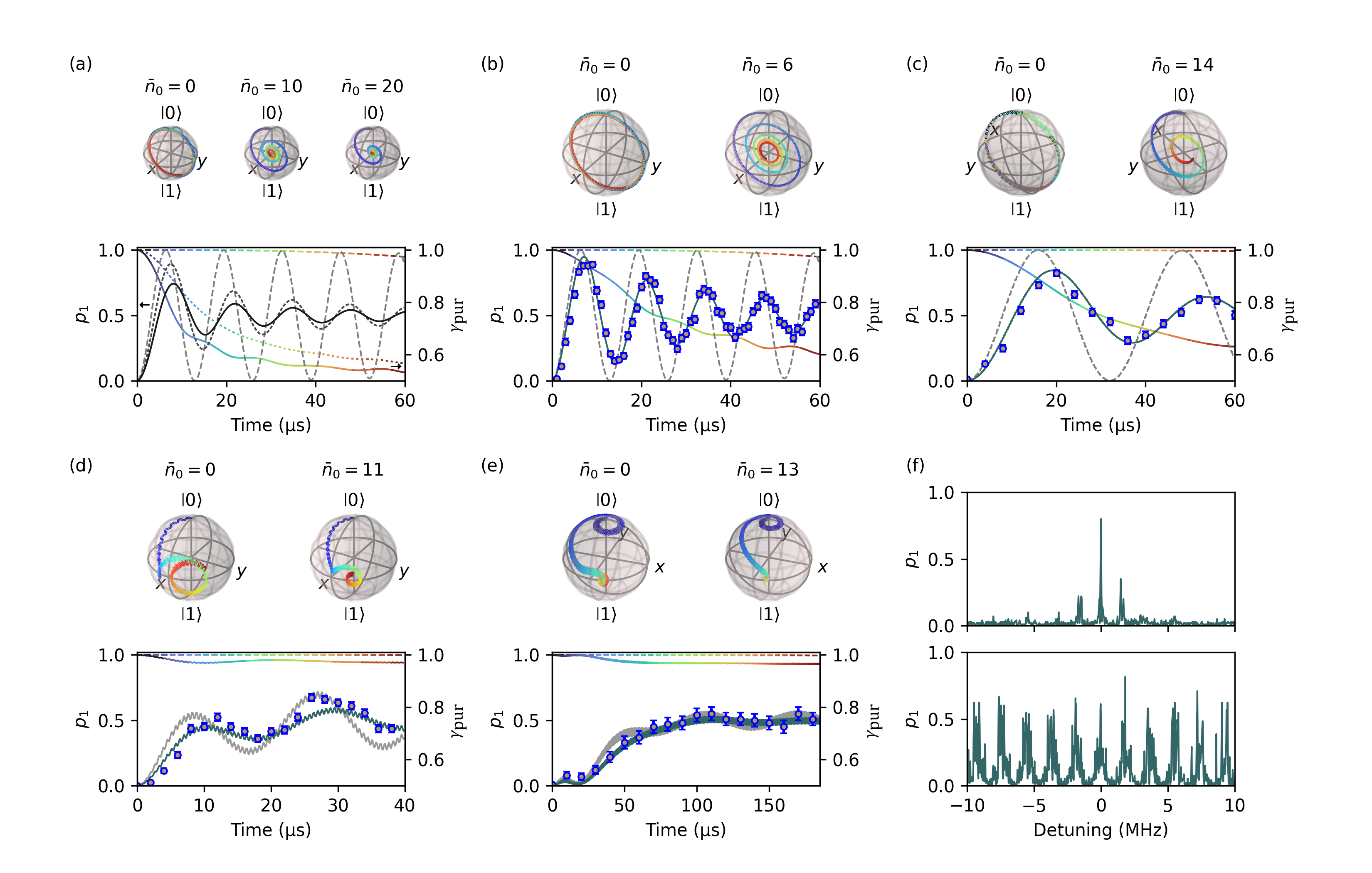}
\caption{Rabi oscillation and Bloch sphere trajectory under the influence of coherent errors induced by semiconductor charging and incoherent errors due to thermal decoherence. (a) Simulated Rabi oscillations for thermal states with the average phonon number $\bar{n}_0=0$ (light grey, dashed), 10 (grey, dotted), and 20 (black, solid). The trajectory on the Bloch sphere and the purity of the qubit state for each case is depicted, with the color map representing the flow of time (blue to red). Traces of the purity and their corresponding Rabi oscillations for identical $\bar{n}_0$'s are represented using the same line styles. (b) Pre-turn-on applied, fully compensated ($t_\mathrm{pre}\gg \tau_\mathrm{str}$, $E_\mathrm{com}=E_\mathrm{str}$). The initial mean phonon number is fitted to $\bar{n}_0=6$. Comparing with the simulation for $\bar{n}_0=0$, the main source of damping occurs from motional decoherence. (c) Pre-turn-on applied, uncompensated ($t_\mathrm{pre}\gg \tau_\mathrm{str}$, $E_\mathrm{com}-E_\mathrm{str}=47\ \mathrm{V}\cdot\mathrm{m}^{-1}$). The initial mean phonon number is fitted to $\bar{n}_0=14$. (d) Pre-turn-on not applied, uncompensated ($t_\mathrm{pre}=0,\ \tau_\mathrm{str}=6$ \textmu s, $E_\mathrm{str}=9\ \mathrm{V}\cdot\mathrm{m}^{-1}$, $E_\mathrm{com}=34\ \mathrm{V}\cdot\mathrm{m}^{-1}$). The initial mean phonon number is fitted to $\bar{n}_0=11$. (e) Pre-turn-on not applied, uncompensated ($t_\mathrm{pre}=0,\ \tau_\mathrm{str}=19$ \textmu s, $E_\mathrm{str}=27\ \mathrm{V}\cdot\mathrm{m}^{-1}$, $E_\mathrm{com}=57\ \mathrm{V}\cdot\mathrm{m}^{-1}$). The initial mean phonon number is fitted to $\bar{n}_0=13$. (f) Sideband spectra depending on the application of the pre-turn-on sequence. The top (bottom) plot is obtained when pre-turn-on is utilized (not utilized). Throughout (b) -- (e), the solid lines are fitted to data, while the 
grey dashed (solid) lines indicate simulation results corresponding $\bar{n}_0=0$ in (b) -- (c) ((d) -- (e)). The purity is represented in solid (dashed) lines for the fitted ($\bar{n}_0=0$) data. The experimental Rabi frequency and heating rate are $\Omega=2\pi\times78$ kHz and $\Gamma\approx10^4$ quanta/s, respectively, for all cases. The simulated ion displacements are on the order of 10 nm -- 100 nm, comparable to the wavelength of the driving laser.}
\label{fig_quantum_sim}
\end{figure*}

In Fig.~\ref{fig_quantum_sim}, experimental data and numerical results are shown for cases with different levels of compensation of the stray fields. Compensation methods include the pre-turn-on sequence and field compensation through dc scans \cite{lee_micromotion_2023}, along with precise alignment of the control lasers to minimize semiconductor charging (see Appendix~\ref{sec_optimization}). The simulated trajectories of the evolution is drawn on the Bloch sphere for each case, and the purity of the state is plotted along with the Rabi oscillation. The purity $\gamma_\mathrm{pur}$ is defined as
\begin{equation}
\label{eqn_purity}
\gamma_\mathrm{pur}=\mathrm{Tr}\left(\rho_\mathrm{qubit}^2\right)
\end{equation}
where $\rho_\mathrm{qubit}$ is the reduced density matrix describing the qubit state of the trapped ion (see Appendix ~\ref{appendix_lindblad}).

The simulation data displayed in Fig.~\ref{fig_quantum_sim} (a) serves as a reference case where there are no background fields and only thermal decoherence is present, for various values of $\bar{n}_0$. Fig.~\ref{fig_quantum_sim} (b) shows the case where both pre-turn-on ($t_\mathrm{pre}\gg \tau_\mathrm{s}$) and complete compensation ($E_\mathrm{c}=E_\mathrm{s}$) have been applied. Resolved sideband cooling has been utilized to cool the ion below the Doppler limit, $\bar{n}_0 \approx 6<10$. Note that the relatively large heating rate of our system ($\Gamma\approx10^4$ quanta/s) prevented us from reaching the motional ground state. Moreover, for the cases where the stray fields were not compensated, corresponding to Fig.~\ref{fig_quantum_sim} (c) -- (e), sideband cooling was completely ineffective.
 
 In Fig.~\ref{fig_quantum_sim} (c), pre-turn-on is applied but the stray field is not compensated. Despite using the same laser parameters as in (b), the observed Rabi frequency is reduced by nearly a factor of 2.5. Note that this reduction cannot be attributed to thermal decoherence. For instance, as shown in Fig.~\ref{fig_quantum_sim} (a), the simulated Rabi frequency is only slightly reduced even with $\bar{n}_0$ = 20, and increasing $\bar{n}_0$ further will result in a decreased visibility before a substantial reduction in the Rabi frequency occurs. On the other hand, our simulations confirm that this phenomenon directly results from the effective modulation of the Raman transition in the presence of excess micromotion \cite{lee_micromotion_2023}, with thermal decoherence mostly being responsible for the decay of the oscillation and purity. Note that the rotational axis of the Rabi oscillation is also modified due to this effective phase modulation.
 
Fig.~\ref{fig_quantum_sim} (d) and (e) show the situations where neither pre-turn-on nor compensation is utilized. The temporal development of the field drastically modifies the Rabi oscillation, giving an impression of substantial thermal decoherence taking place. Again, we emphasize that such patterns in the Rabi oscillation cannot be reproduced by merely adding more phonons or increasing the heating rate. Instead, simulation results show that the apparent decay in the Rabi oscillation is actually a consequence of coherent rotational errors in the Bloch sphere rather than incoherent dephasing or damping. Indeed, these errors are pronounced even when the ion is in its motional ground state ($\bar{n}_0=0$). Interpretation of the Bloch sphere trajectory is given as follows. The motion of the ion causes a continuous transformation of the rotational axis of the Rabi oscillation, through the time-dependent spatial phase $\Delta k x(t)$. Once $F(t)$ in Eq.~(\ref{eqn_field}) converges to $F(t\gg\tau_\mathrm{str})=e(E_\mathrm{str}-E_\mathrm{com})$, Rabi oscillation proceeds around the axis on the equator of the Bloch sphere, as determined by this value.

\begin{figure}[h]
\centering
\includegraphics[width=1\columnwidth]{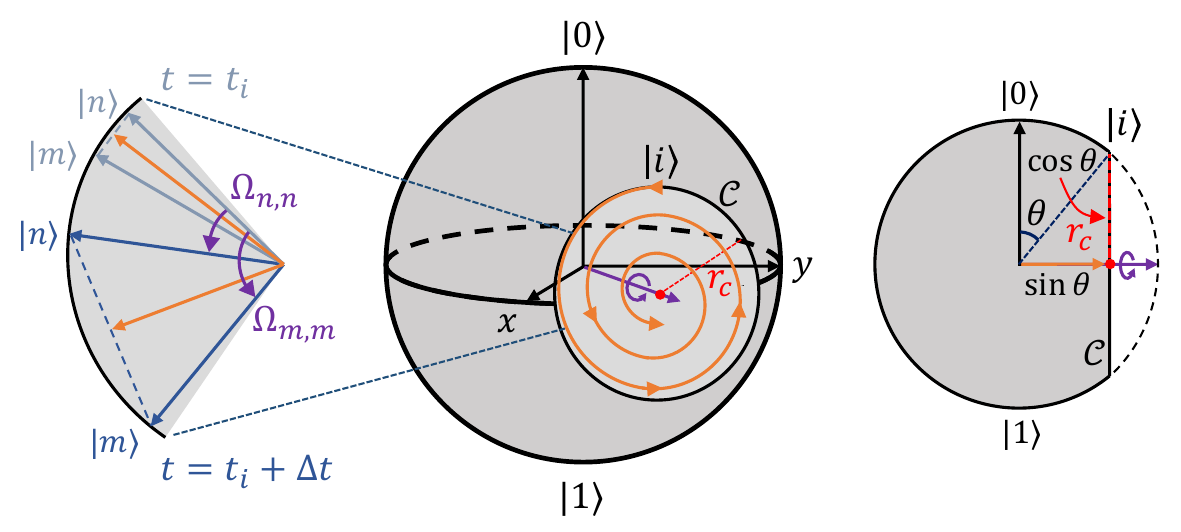}
\caption{Geometric interpretation of thermal decoherence on the Bloch sphere. An initial state $\ket{i}$ that undergoes thermally damped Rabi oscillation around a rotational axis on the equator of the Bloch sphere will spiral in towards the center of the circle $\mathscr{C}$ with radius $r_{c}$ (the rotational axis, quantum state, and radius $r_{c}$ are colored in purple, orange, and red, respectively). This is because components of the state with different Rabi frequencies originating from the phonon distribution will spread, adding up incoherently during the evolution. This process is schematically depicted in the left figure, where carrier Rabi frequencies for the oscillator states $\ket{n}$ and $\ket{m}$ are denoted as $\Omega_{n,n}$ and $\Omega_{m,m}$, respectively. The vector in orange represents the quantum state losing purity within a time interval $\Delta t$. The extent of the spread is proportional to the radius $r_{c}=\mathrm{cos}\theta$, thus, resulting in a greater loss of purity as $r_{c}$ increases. In the right figure, the vector in orange represents the qubit state with the lowest purity given by $\gamma_\mathrm{pur,min}=0.5+\mathrm{sin}\theta$, assuming $\ket{i}$ and the specified rotational axis.}
\label{fig_bloch}
\end{figure}

Despite the distinct Rabi oscillation patterns shown in Fig.~\ref{fig_quantum_sim} (d) and (e), in both cases, the purity only decreases during the rise time of the stray field ($\tau_\mathrm{str}$), and shows little sign of decoherence afterwards. To understand this, it is helpful to picture the thermally damped Rabi oscillation on the Bloch sphere. In Fig.~\ref{fig_bloch}, a cross-sectional view of the Bloch sphere is shown where an initial state $\ket{i}$ undergoes within Rabi oscillation around the rotation axis (purple). The circle $\mathscr{C}$ is the trajectory that an ideal coherent quantum state would traverse at the Rabi frequency. In the presence of thermal decoherence, however, the state (orange) will spiral into the circle $\mathscr{C}$, eventually reaching the center of the circle. The state becomes a mixed state as it evolves into the interior of the Bloch sphere. The purity, however, does not simply fall to 0.5, but has a lower bound given by $\gamma_\mathrm{pur, min}=0.5(1+\mathrm{sin}\theta)$. The geometrical representation for the term $\mathrm{sin}\theta$ is provided in Fig.~\ref{fig_bloch}. The larger the radius of the circle $\mathscr{C}$, $r_{c}=\mathrm{cos}\theta$ (red), the larger the decoherence, with $\gamma_\mathrm{pur, min}=0.5$ for $\ket{i}=\ket{0}$ ($\theta=0$). This is because the incoherent sum of Rabi frequencies arising to the phonon distribution is largest when the radius $r_{c}$ is at its maximum (see the captions of Fig.~\ref{fig_bloch}).

This explains the traces of the purity simulated in Fig.~\ref{fig_quantum_sim} (d) and (e). In both cases, the purity is significantly reduced as the state $\ket{i}=\ket{0}$ is initially rotated about an axis on the equator of the Bloch sphere, due to the maximal radius $r_{c}=1 \ (\theta=0)$. During the rise time of the stray field, the rotational axis transforms as well, eventually reaching a new orientation on the equator. If the state is near the new rotational axis by the time that $F(t)$ converges, the subsequent trajectory will be confined within a smaller circle $\mathscr{C}$. This prevents the qubit state from moving further into the interior of the Bloch sphere. Ironically, although their Rabi oscillations seems to occur in an incoherent manner, the states are actually more coherent than those in Fig.~\ref{fig_quantum_sim} (b) and (c), where the dynamical effects of the stray field have been compensated. These observations demonstrate that photo-induced stray fields and excess micromotion mainly induce coherent errors, and do not necessarily increase thermal decoherence.

The effect of the pre-turn-on sequence on the sideband spectra is shown in Fig.~\ref{fig_quantum_sim} (e). When the evolution is initiated only after the stray field reaches its steady state, the sideband spectrum exhibits just the carrier and secular motion transition peaks. On the other hand, when the pre-turn-on sequence is not applied, we observe numerous transition peaks due to the convolution of the ion oscillation at the motional frequency with the trap frequency and the temporal profile of the stray field (see Eq.~(\ref{eqn_alpha})). Again, such a spectrum cannot be obtained by merely heating up the ion as only transitions at higher order motional frequencies will occur with a decaying profile.

\section{Discussion}
\label{sec_discussion}

Our model provides insight into previous studies on dielectric charging \cite{harlander_trapped-ion_2010, wang_laser-induced_2011, harter_long-term_2014, doret_controlling_2012, hong_new_2017, ivory_integrated_2021, jung_2023} and reported issues on silicon charging \cite{niedermayr_cryogenic_2014, lakhmanskiy_heating_2019, mehta_integrated_2020}. The conventional explanation for the charging of insulators in ion traps is attributed to the photoelectric effect, encompassing direct photoemission from insulators or the capturing of photoelectrons emitted from interfacing conductors into insulators \cite{harlander_trapped-ion_2010}. On the other hand, internal carrier dynamics or phenomena that arise due to boundaries or inhomogeneities have largely been neglected. According to our study, such factors can have significant impact in the overall response of the system. Here, we enumerate potential mechanisms that can enhance our understanding of dielectric charging, aiming to shed light on previously unexplained observations, such as the wide range of time scales manifest in the relaxation process and irregularities in the spectral response \cite{harlander_trapped-ion_2010, wang_laser-induced_2011, harter_long-term_2014, ivory_integrated_2021, jung_2023}.

Considering that the lasers commonly used in ion traps operate within an energy range of 1 -- 4 eV, and typical solids have work functions ranging from 4 -- 6 eV \cite{stanislaw_1998} (also see Table~\ref{tbl_material} where insulators are characterized by the bandgap energy $E_\mathrm{g}$ and electron affinity $\chi$, while conductors are specified by the work function $\phi_\mathrm{m}$, respectively), external photoemission should mainly occur through nonlinear processes involving multiphoton absorption. Note, there are exceptional cases in which a linear photoelectric response is observed below 4 eV from aluminum \cite{baikie_2014}, or when certain lasers used for photoionization exceed 5 eV, such as in $^{9}$Be$^+$ systems. In photoemission spectroscopy studies, multiphoton photoemission has been primarily demonstrated for conductors where the free carrier concentration is large, mostly using pulsed lasers \cite{bechtel_1977, damascelli_1996}, and rarely with continuous-wave lasers \cite{sivis_2018}. However, this has been accepted as the main cause for dielectric charging in ion trap systems, even with weak continuous lasers, due to the extreme sensitivity of ions to sense sources as small as 10 -- 1000 elementary charges and the enhancement of charging correlated with UV light. Nevertheless, this assumption is simplified and requires careful examination.

Let us consider the dielectric relaxation times $\tau_+$ of commonly used insulators in ion trap chips listed in Table~\ref{tbl_material} (see Appendix~\ref{appendix_bulk_2}). They typically fall within the order of hours, which is compatible with long-term charging measured in Ref.~\cite{harter_long-term_2014}, but distinct to transient responses on the order of 10 -- 100 s reported in Refs.~\cite{harlander_trapped-ion_2010, ivory_integrated_2021, jung_2023}. Characteristic time scales significantly distinct from the dielectric relaxation time indicate the existence of underlying carrier dynamics in the presence of inhomogeneities throughout the bulk and boundaries (see Appendix~\ref{appendix_bulk_2}). 

Specifically, internal carrier dynamics induced by linear or nonlinear absorption of light in such solids \cite{brandi_1983, desalvo_1996, tanaka_2003}, associated with microscopic bond and defect structures \cite{salh_2011, linards_2021} has been largely overlooked in the context of dielectric charging. Considering that photo-induced response from localized electronic states in silica has been observed, where the associated time scales are on the order of 10 -- 100 s \cite{trukhin_2019}, it may not be reasonable to neglect this phenomena. Moreover, even photoemission involves carrier dynamics as energetic electrons experience inelastic scattering as a function of kinetic energy during their transport to the surface \cite{seah_1979}, in conjunction with the relaxation and recombination of holes.

\begin{table}[h]
\caption{Properties of commonly occurring insulators and conductors in ion trap chips}
\begin{tabular}{cccccc}
\hline
                       & \multicolumn{3}{c}{Insulator} & \multicolumn{2}{c}{Conductor} \\ \hline
Material               &  $\mathrm{SiO}_{2}$    & $\mathrm{Al}_{2}\mathrm{O}_{3}$    & $\mathrm{Si}_{3}\mathrm{N}_{4}$    & Cu            & Au            \\ \hline
$E_\mathrm{g}$ {[}eV{]} & 9       & 7      & 5.3        & --         & --          \\ \hline
$\chi, \ \phi_\mathrm{m}$ {[}eV{]}       & 1.1     & 2        & 2.1      & 4.6          & 5.5          \\ \hline
$\epsilon$             & 3.9     &  9     & 7      & --           & --           \\ \hline
$\sigma \ [\mathrm{\Omega^{-1}}\cdot s^{-1}]$                & $10^{-15}$        & $10^{-14}$         &  $10^{-14}$        & $10^7$              & $10^7$              \\ \hline
$\tau_{+}$             & 9.6 h  & 2.2 h  & 1.7 h      & $10^{-15}$ s      &  $10^{-15}$ s             \\ \hline
References             & \cite{cook_SiO_2003, robertson_di_2004}     & \cite{robertson_di_2004, afanasev_AlO_2011, filatova_AlO_2015}      & \cite{cook_SiN_2003, robertson_di_2004}      &  {\cite{mitchell_copper_1951, johnson_copper_1975}}    & {\cite{matula_gold_1979, olmon_gold_2012, ishida_2020}}              \\ \hline
\end{tabular}
\label{tbl_material}
\end{table}

\begin{table*}
\caption{Categorization of stray electric fields in ion trap chips.}
\centering
\begin{tblr}{
  width = \linewidth,
  colspec = {Q[90]Q[130]Q[150]Q[190]Q[190]Q[170]},
  cells = {c},
  cell{1}{2} = {c=3}{0.2\linewidth},
  cell{1}{5} = {c=2}{0.3\linewidth},
  cell{5}{2} = {c=3}{0.3\linewidth},
  cell{5}{5} = {c=2}{0.33\linewidth},
  %vlines,
  hline{1,3-7} = {-}{},
  hline{2} = {2-4}{},
  hline{2} = {5-6}{},
}
           & \textbf{Electric-field noise}     &                                      &                                                                 & \textbf{Photo-induced electric field}         &             \\
           & \textbf{Electrode noise} & \textbf{Thin layer on metal}        & \textbf{Patch potential, two-level fluctuators, adatoms, ... }         & {\textbf{Dielectric} \\ \textbf{charging}}            & \textbf{Semiconductor charging}\\
{Mechanisms}  & Noise
  from resistance    & Thermal noise from dissipation         & Phonon-induced fluctuation
  of charges   & {Photoemission \\ Charge capture \\ Internal dynamics}   & {
  Surface photovoltage
  \\Bulk response \\(Dember effect)}\\
Material   & Electrode conductor or semiconductor
         & Insulating layer on conductor          & Conductor surface (also semiconductor for two-level fluctuators)                                            & {Insulator \\ (including boundaries)} & Semiconductor (including boundaries)\\
{Effect\\ on ion}     & {Heating, motional dephasing, \\ incoherent errors} & & & {Displacement, \\ coherent errors} \\
References & Turchette 2000 \cite{turchette_heating_2000}
  & {Kumph 2016 \cite{kumph_electric-field_2016} \\Boldin 2018 \cite{boldin_measuring_2018}} & {Turchette 2000 \cite{turchette_heating_2000} \\Schriefl 2006 \cite{schriefl_decoherence_2006}
  \\Safavi-Naini 2011 \cite{safavi-naini_microscopic_2011} \\Boldin 2018 \cite{boldin_measuring_2018} }
  & {Harlander 2010 \cite{harlander_trapped-ion_2010} \\Wang 2011 \cite{wang_laser-induced_2011} \\ H\"arter 2014 \cite{harter_long-term_2014} \\ Ivory 2021 \cite{ivory_integrated_2021}} 
  & {This work     \\ Lakhmanskiy 2020 \cite{lakhmanskiy_heating_2019}  \\ Mehta 2020 \cite{mehta_integrated_2020}}
\end{tblr}
\label{tbl_category}
\end{table*}

Boundaries in a system, such as the surfaces or interfaces between materials, are particularly important because defects and barriers form at the boundaries \cite{yeo_2002}, which can introduce various types of carrier emission and capturing mechanisms \cite{williams_1965, lauer_2003, afanas_2019}. Indeed, certain reports on dielectric charging pertain to insulator/conductor structures \cite{harlander_trapped-ion_2010, wang_laser-induced_2011, jung_2023}, where the net effect actually arises as a joint process involving the two materials. In fact, multiphoton photoemission has been utilized to investigate surface states on conductors \cite{giesen_1985}, interface states of insulator/conductor structures \cite{padowitz_1992, martin_2002, rohleder_2005, gillmeister_2018, reutzel_2020}, and transfer of electrons from conductors into insulators \cite{james_2010}. These are all probable microscopic processes that have been grouped into a single mechanism in the context of dielectric charging. 

In contrast to the relatively fast response time of our silicon substrate, a slower charging process on the order of 1 -- 10 s has been reported in a cryogenic silicon-based chip \cite{lakhmanskiy_heating_2019}. This is intriguing because a chip fabricated earlier under similar conditions showed no such issues \cite{niedermayr_cryogenic_2014}. This implies that fabrication conditions have significant impact on the surface quality of the chip. Both chips used the deep RIE procedure, which is suspected to be the cause of the defect states on the surface of our substrate. A difference between our chip and the one discussed in Ref.~\cite{lakhmanskiy_heating_2019} is that the latter operates at cryogenic temperatures, utilizes intrinsic silicon, and creates a thermal dioxide layer on the silicon surface, which presumably results in different surface conditions that are responsible for the different characteristic time scale.

Based on the simulation results of our semiconductor charging model, we list some implications for the development of semiconductor-based ion trap chip fabrication. First, increasing the bandgap or bulk doping concentration may not necessarily reduce SPV effects significantly as long as the influence of surface defects or interface states cannot be controlled. SPV can be drastically reduced only when the substrate behaves like an insulator (low free carrier density, low mobility) or a conductor (high free carrier density, high mobility) \cite{luth_sem_2012}.

Second, the charging mechanism introduced in our study is not likely to disappear by merely changing the substrate to n-type silicon. For example, if the surface is oxidized so that the electron density is high at the surface in thermal equilibrium (accumulation layer), optical excitation from defect states into the conduction band may be suppressed. Even if this is the case, there can exist more defect states within the bandgap due to the elevated Fermi level. The net effect of these mechanisms must be scrutinized carefully in order to predict the resultant SPV. 

Third, decreasing the substrate temperature is not necessarily beneficial unless the temperature is lowered to sub-Kelvin levels. This is because the diffusion of carriers, which is proportional to the product of the  temperature and carrier mobility (see Appendix~\ref{appendix_sem_eq}), may not be greatly reduced as the mobility actually increases by orders of magnitude \cite{reggiani_2000}. Also, even when the temperature of the surface is substantially low, generation-recombination noise at the illuminated surface may lead to residual heating of the ion. 

Finally, while techniques like surface passivation can be employed to mitigate unwanted surface states \cite{meiners_passivation_1988, mizsei_passivation_2002}, this is not always feasible. For ion trap systems, it seems best to optically block the exposed surfaces completely using reflective metal layers \cite{blain_hybrid_2021}.

A categorization of the stray fields that have been reported in ion traps is presented in Table~\ref{tbl_category}. Our study is summarized in the column for semiconductor charging. As mentioned in Sec.~\ref{sec_intro}, field noise mainly contributes to ion heating. The reported heating rates of the ion in microfabricated ion traps so far mostly follow the distance scaling of $\sim d^{-4}$, frequency scaling of $\sim \omega^{-2}$ and temperature scaling of $\sim T^{1.2}$ \cite{auchter_industrially_2022, sedlacek_distance_2018}, but there exists some inconsistency in their absolute levels. When comparing heating rates measured in silicon-based traps and glass-based traps \cite{auchter_industrially_2022, boldin_measuring_2018, blain_hybrid_2021, an_distance_2019, sedlacek_distance_2018, holz_2d_2020, spivey_high-stability_2021, niedermayr_cryogenic_2014, mehta_ion_2014, lakhmanskiy_heating_2019}, the latter typically appears to reach lower heating rates for all relevant scaling factors, although the trend is not perfectly clear. We speculate that the difference may be partially ascribed to an unexplored aspect of photo-induced charging, involving the fluctuation or generation-recombination noise of unpaired/excess charges, which may act as additional noise sources. It remains intriguing to validate this conjecture through a more controlled measurement assessing the dependence of the heating rate on the substrate material.

\section{Conclusion}
\label{sec_conclusion}

We have observed and analyzed the photo-induced charging process of the silicon substrate in a microfabricated chip by direct measurement of the stray field through motion-sensitive transitions of a trapped ion. A semiconductor charging model based on the SPV theory has been presented. The dominant charging mechanism is identified as SPV inversion in silicon, which occurs irrespective of incident wavelength, primarily attributed to surface defects introduced during the microfabrication process. We have characterized the stray field in multiple ways, including direct imaging, measurement of micromotion-modified transition probability \cite{lee_micromotion_2023}, and the time-resolved Doppler shift measurement. 
Analysis of motion-sensitive qubit transitions revealed that coherent errors are induced by stray fields, which could be mitigated using well-tuned control procedures. Finally, the implications of our model with respect to other photo-induced charging mechanisms and the fabrication of semiconductor-based chips have been discussed. Limitations of our semiconductor charging model and possible alternatives are discussed in Appendix \ref{appendix_lim}.

\section*{Acknowledgement}
This work was supported by the Institute for Information \& communications Technology Planning \& Evaluation (IITP) grant (No. 2022-0-01040), the National Research Foundation of Korea (NRF) grant (No. 2020R1A2C3005689, No. 2020M3E4A1079867), and the Samsung Research Funding \& Incubation Center of Samsung Electronics (No. SRFC-IT1901-09).

\medskip

%\section{Author contributions}
W.L. and D.C. developed the theoretical model and performed the experiments. D.C. selected theories from the literature and carried out the numerical simulations. C.K. examined and refined the theory. W.L and H.J constructed the Raman laser setup and made a first observation. B.C. conducted chip simulation. K.Y. and S.Y. conducted test fabrication. C.J., J.J., D.D.C. and T.K were involved in chip fabrication and setup construction. T.K. supervised the project. W.L., D.C., and T.K. conceived the project and wrote the paper. All authors discussed the results and commented on the manuscript.

%\section{Competing interests}
%The authors declare no competing interests.

\appendix

\section{Experimental system}
\label{sec_system}

A schematic cross-sectional illustration of the trap chip and incident laser beams is shown in Fig.~\ref{fig_chip}. (More detailed descriptions of the chip architecture can be found in Ref. \cite{jung_microfabricated_2021}.) The trap chip was fabricated on a silicon substrate which is boron-doped with a concentration of $10^{15}$ cm$^{-3}$, through MEMS technology. The electrodes are made of aluminum alloy with 1\% copper and they are extended to the sidewalls of the underneath pillars to prevent the charging effect of dielectrics induced by lasers. The electrodes near the trapping region are additionally coated by gold to avoid oxidation. There is also a loading slot with a width of 80 \textmu m in the middle of the trap chip running along the trap axis ($\hat{z}$) direction, originally made for the purpose of backside loading of atoms.

$^{171}$Yb$^+$ ions are trapped on the chip at a height of 80 \textmu m in an ultrahigh vacuum of $< 1\times10^{-10}$ mbar. A trapped ion is tightly confined along the transverse directions ($\hat{x}$, $\hat{y}$) in a pseudo-potential generated by rf voltages with a frequency of 22.21 MHz, and loosely confined along the trap axial direction ($\hat{z}$) in a static potential generated by a set of dc voltages. The trap secular frequencies are 1.6 MHz, 1.5 MHz, and 450 kHz for the three principal axes.

A 369-nm cooling beam and a 935-nm repumping beam are injected in a counter-propagating configuration, 45$^{\circ}$ to the trap axis and parallel to the trap chip surface ($\hat{x} + \hat{z}$). The powers of these lasers are 3 \textmu W and 30 \textmu W, and the beam waists of the lasers at the ion position are 15 \textmu m and 45 \textmu m, respectively. The fluorescence of the trapped ion is imaged by a high-NA imaging lens (Photon Gear 15470-S, NA 0.6) and detected by an EMCCD or a photomultiplier tube (PMT).

For Raman transition between $\ket{0}=\ket{^2 S_{1/2}, F=0, m_F = 0}$ and $\ket{1}=\ket{^2 S_{1/2}, F=1, m_F = 1}$, a 355-nm picosecond pulse laser with a repetition rate of 120.127 MHz is split into two beams and separately modulated with AOM's to become a pair of beatnote-locked Raman beams (beam 1 and 2, or beam 1 and 3), for control of the qubit and motional states of the ions. Raman beam 1 is assigned for individual addressing of ions, so is tightly focused by the imaging lens to a waist of 2 \textmu m and is directed to the ions in a direction perpendicular to the chip. Raman beam 2 (or 3) is assigned for global addressing of the entire ion chain and had two alternative configurations. For the diagnosis of the laser-induced field in out-of-plane direction to the trap chip as described in Sec.~\ref{sec_null_shift}, and for later mitigation of laser-induced field, the global beam (Raman beam 2) was incident from the backside of the trap chip, in counter-propagating configuration with Raman beam 1, with a waist of 15 \textmu m. On the other hand, for the measurement of frequency shift described in Sec.~\ref{sec_frequency_shift}, the global beam (Raman beam 3) was incident on the ion in a direction perpendicular both to the trap axis and to Raman beam 1 ($\hat{x}$). The waist was 41 \textmu m along the trap axis direction and 26 \textmu m along the out-of-plane direction.

The alternative 935-nm probe laser used for measurement in Sec.~\ref{sec_null_shift} was vertically injected from the backside of the chip to penetrate through the loading slot with a diameter of 60 \textmu m. The intensity of the laser beam at the ion position was fixed at around 50 mW/cm$^2$.

\section{Scheme for optimization of quantum control under effects of semiconductor charging}
\label{sec_optimization}

Several treatments have been applied to suppress the effect of the photo-induced stray field. The first was minimization of the excess micromotion, by means of the described compensation voltage measurement. The maximum Rabi frequency of the Raman transition guarantees that the excess micromotion was truly minimized. Second, we employed the pre-turn-on scheme, as described in  Sec.~\ref{sec_frequency_shift}, in our actual quantum control sequences as well. One of the beams, which causes the largest stray field (usually the global Raman beam), was turned on tens of milliseconds prior to the control of the qubit. This mitigates the rapid drift of the resonance frequency at laser turn-on, by delaying the actual evolution from the initial transient drift. This reduced the change in the frequency of the ion qubit and improved its coherence, although not to a sufficiently high level, probably due to abundance of thermal charge carriers generated and heated by long irradiation. Third, the direction of the global Raman laser injection was changed from side-injection (perpendicular) to back-injection (counter-propagating), as we discovered it improves the coherence (increase in the peak probability of the Rabi oscillation by around 0.1). Additionally, the alignment was further optimized to minimize the stray field effect in the quantum control as much as possible. The Raman beams passing through the chip slot were kept as far as possible from the both inner sides of the substrate, and exactly perpendicular to the chip surface. The global Raman beam from the backside was aligned with this criterion by imaging the beam when they are at the edges of the chip slot and then positioning the beam at the exact center of the slot, while maintaining the maximum Rabi frequency.

\section{Theoretical background of the photoconductive charging model}
\label{appendix_model}
The dynamics of the charge density and potentials in semiconductors is completely described by simultaneously solving the continuity and Poisson equations, also known as the semiconductor equations \cite{roosbroeck_sem_1950}. Obtaining analytical solutions to these equations is a formidable task due to the highly coupled nature of the equations and nonlinearity present in numerous terms.

Three types of approximations are often applied to circumvent this problem \cite{roos_sem_1979, hout_sem_1996}. The first is to limit the analysis to doped or extrinsic semiconductors which partially decouples the equations, i.e., the minority and majority carrier equations. The second is to consider the low excitation regime where the system is not very far from thermal equilibrium so that nonlinear terms are negligible. Finally, local charge neutrality or quasi-neutrality is assumed in order to fully linearize and decouple the equations. As will be explained in the following section, local charge neutrality, despite its practicality in limiting cases, is problematic in general situations. Therefore, it will be replaced by the global charge neutrality condition \cite{krcmar_exact_2002}, which is the physically correct constraint with respect to total charge conservation. Note that external static fields and lattice heating effects are assumed to be negligible throughout the analysis.

We consider a semiconductor slab of thickness $l$ across whose surfaces ($x=0,\ l$) flow of charge carriers is inhibited. Light is shone on surface $x=0$ while surface $x=l$ is electrically grounded. The photoconductive charging model can be classified into two cases depending on 1) the absence of surface charges (the uniform bulk) and 2) the presence of surface charges. It is important to understand 1) because the bulk response of an illuminated semiconductor contains valuable information about the natural dynamics of carriers in non-equilibrium. Analytical solutions for the full spatiotemporal structure of the charge density and potential can be obtained by using only the low excitation regime approximation. In real semiconductor surfaces, 2) is usually the dominant source of photovoltage, whose exact treatment is often challenging and hence requires a numerical approach.

\section{Semiconductor equations without local charge neutrality}
\label{appendix_sem_eq}
The semiconductor equations describe the dynamics of three quantities: the electron (n) and hole (p) densities, and the electrostatic potential $\phi$. We use a dimensionless quantity, $u=\beta \phi$, where $\beta^{-1}=kT/e$ is the thermal energy evaluated in volts. It can be interpreted as the potential evaluated in units of $\beta^{-1}$ or equivalently, as the energy measured in units of $kT$. The electron and hole carrier flux, $j_\mathrm{n}$ and  $j_\mathrm{p}$, are defined through the relations
\begin{equation}
\label{eqn_a1}
\frac{j_\mathrm{n}}{D_\mathrm{n}} =-\frac{\partial n}{\partial x} + n \frac{\partial u}{\partial x},\   \frac{j_\mathrm{p}}{D_\mathrm{p}} =-\frac{\partial p}{\partial x} - p \frac{\partial u}{\partial x}
\end{equation}
Here, $D_\mathrm{n}=\mu_\mathrm{n} \beta^{-1},\ D_\mathrm{p}=\mu_\mathrm{p} \beta^{-1}$ are the diffusion coefficients where $\mu_\mathrm{n},\mu_\mathrm{p}$ are the carrier mobilities. We adopt definitions for the carrier densities from the references \cite{garrett_physical_1955, johnson_large-signal_1958},
\begin{equation}
\label{eqn_a2}
n=n_0+\delta n=n_\mathrm{i} e^{u-u_{\mathrm{F_n}}},\ p=p_0+\delta p=n_\mathrm{i} e^{u_{\mathrm{F_p}}-u}
\end{equation}
where $n_0,\ p_0$ are the carrier densities in thermal equilibrium, $\delta n,\ \delta p$ the excess carrier densities in non-equilibrium, and $n_i$ is the intrinsic carrier concentration. We have used $u_\mathrm{F_n}=\beta \phi_\mathrm{F_n},\ u_\mathrm{F_p}=\beta \phi_\mathrm{F_p}$ where $\phi_\mathrm{F_n },\ \phi_\mathrm{F_p}$ are the quasi-Fermi potentials of electrons and holes. Unless stated otherwise, the subscript 0 stands for a quantity evaluated in thermal equilibrium ($\delta n=\delta p=0$), where the equilibrium temperature is assumed as $T$=300 K. Note that $u_\mathrm{F_n,0}=u_\mathrm{F_p,0}=u_F=\beta \phi_\mathrm{F}$ where $\phi_\mathrm{F}$ is the Fermi potential of the semiconductor that is determined by the bulk doping concentration which is assumed to be uniform. This implies the following expressions for $u_\mathrm{F_n},\ u_\mathrm{F_p}$, which can in turn be interpreted as their definitions
\begin{equation}
\begin{aligned}
\label{eqn_a3}
u_\mathrm{F_n}=u_\mathrm{F}+\delta u-\mathrm{ln}⁡\left( 1+ \frac{\delta n}{n_0} \right) \\ u_\mathrm{F_p}=u_\mathrm{F}+\delta u+\mathrm{ln}⁡\left( 1+\frac{\delta p}{p_0} \right)
\end{aligned}
\end{equation}

\begin{figure}[h]
\centering
\includegraphics[width=0.65\columnwidth]{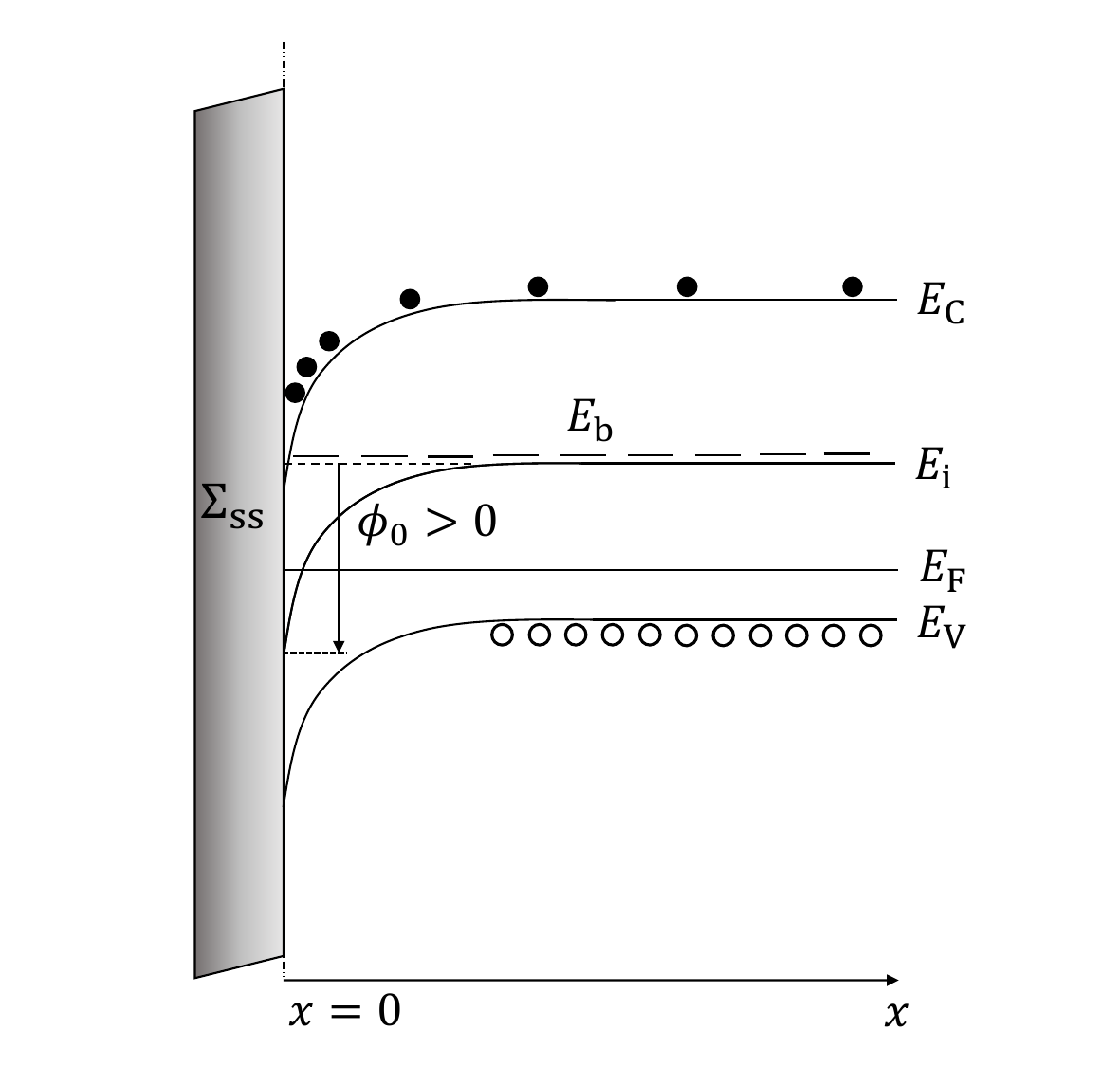}
\caption{Band diagram of a semiconductor in thermal equilibrium. Note that the energy level is defined as $E=-\beta \phi$ for the corresponding potentials $\phi$. $E_\mathrm{C}$ and $E_\mathrm{V}$ indicate the conduction and valence band edges.}
\label{fig_model_appendix}
\end{figure}

The quantity $\delta u=u-u_0=\beta(\phi -\phi_0 )$ represents the difference between potentials in non-equilibrium and thermal equilibrium. Conversely, $u=u_0+\delta u$. We will call $u_0 \ (\delta u)$ the equilibrium (excess) potential. A graphical representation of the potentials is provided in the energy band diagrams in Fig.~\ref{fig_model_appendix}. Scaling the intrinsic Fermi potential to zero, $u_\mathrm{F_i}=\beta \phi_\mathrm{F_i}=0$, the sign convention is that the value of a potential is positive when it lies below $u_\mathrm{F_i}=0$, and negative when it is above. Provided the definitions listed above, the set of semiconductor equations is obtained as \cite{roosbroeck_sem_1950, kingston_calculation_1955}

\begin{widetext}
%\begin{equation}
\begin{align*}
\label{eqn_a4}
-\frac{1}{D_\mathrm{n}} \frac{\partial n}{\partial t}&=\frac{\partial}{\partial x} \left(\frac{j_\mathrm{n}}{D_\mathrm{n}}\right)+\frac{R_\mathrm{b}-G_\mathrm{b}}{D_\mathrm{n}} \rightarrow   \frac{1}{D_\mathrm{n}}   \frac{\partial n}{\partial t}=\frac{\partial^2 \delta n}{\partial x^2}-\frac{\partial u}{\partial x} \frac{\partial \delta n}{\partial x}-\frac{\partial^2 u}{\partial x^2} \delta n -n_0\left(\frac{\partial u_0}{\partial x} \frac{\partial \delta u}{\partial x} + \frac{\partial^2 \delta u}{\partial x^2} \right)-\frac{R_\mathrm{b}-G_\mathrm{b}}{D_\mathrm{n}} \ \ \ \ \
\end{align*}
\begin{align}
-\frac{1}{D_\mathrm{p}} \frac{\partial p}{\partial t}&=\frac{\partial}{\partial x} \left(\frac{j_\mathrm{p}}{D_\mathrm{p}}\right)+\frac{R_\mathrm{b}-G_\mathrm{b}}{D_\mathrm{p}} \rightarrow   \frac{1}{D_\mathrm{p}}   \frac{\partial p}{\partial t}=\frac{\partial^2 \delta p}{\partial x^2} + \frac{\partial u}{\partial x} \frac{\partial \delta p}{\partial x}+\frac{\partial^2 u}{\partial x^2} \delta p -p_0\left(\frac{\partial u_0}{\partial x} \frac{\partial \delta u}{\partial x} - \frac{\partial^2 \delta u}{\partial x^2} \right)-\frac{R_\mathrm{b}-G_\mathrm{b}}{D_\mathrm{p}}
\end{align}
\begin{align*}
\frac{\partial^2 u}{\partial x^2}&=-\beta \frac{e}{\epsilon_0 \epsilon} (p-n-p_\mathrm{b}+n_\mathrm{b} )\rightarrow 
\left\{\begin{matrix}
\frac{\partial^2 u_0}{\partial x^2}=\frac{2}{\lambda_\mathrm{D_i}^2} [\mathrm{sinh}⁡(u_0-u_\mathrm{F})+\mathrm{sinh}⁡(u_\mathrm{F})] \\ \frac{\partial^2 \delta u}{\partial x^2}=-\beta \frac{e}{\epsilon_0 \epsilon} (\delta p - \delta n)= -\frac{1}{\lambda_\mathrm{D_X}^2}  \frac{\delta p - \delta n}{n_\mathrm{X_b}} \end{matrix} \right.
\end{align*}
%\end{equation}
\end{widetext}
where $R_\mathrm{b}$ and $G_\mathrm{b}$ are the net recombination and generation rates of charge carriers occurring within the bulk, $0<x<l$. $\epsilon_0,\ \epsilon$ are the permittivity of free space and the dielectric constant of the semiconductor, and $n_b=n_\mathrm{i} e^{-u_\mathrm{F}},\ p_b=n_\mathrm{i} e^{u_\mathrm{F}}$ the carrier densities in thermal equilibrium when $u_0=0$. Finally, $\lambda_\mathrm{D_i}=\left(\epsilon_0 \epsilon/\beta en_\mathrm{i}\right)^{1/2}$ is the intrinsic Debye length, and $\lambda_\mathrm{D_X}=\left(\epsilon_0 \epsilon/\beta en_\mathrm{X_b}\right)^{1/2}$ the Debye length associated with density $n_\mathrm{X_b}$. We can use $n_\mathrm{X_b}=n_\mathrm{i}$ for an intrinsic type or $n_\mathrm{X_b}=p_\mathrm{b} \ (n_\mathrm{b} )$ for extrinsic p-type (n-type) semiconductors.

The local charge neutrality assumption identifies the excess electron and hole densities throughout the body of the semiconductor ($\delta n=\delta p$), hence, $\partial^2 \delta u/\partial x^2=0$. Since the total charge density is defined as $\rho=\delta p- \delta n$, this amounts to removing the Poisson equation from the semiconductor equations, and nulling any effects occurring from the total charge distribution. This hinders one from evaluating the exact solution for $\delta u$ which is the focus of our study.

Therefore, the global charge neutrality condition, which states that the total charge be conserved within the semiconductor as a whole, is introduced. In order to express this statement quantitatively, we present the appropriate boundary conditions for the free surfaces of a semiconductor. The boundary conditions at the surfaces are determined by the charge carrier flux across the surfaces
\begin{equation}
\begin{gathered}
\label{eqn_a5}
\frac{j_\mathrm{n}}{-D_\mathrm{n}} \Bigr|_{x=0,l}=\left(\frac{\partial \delta n}{\partial x}-\frac{\partial u_0}{\partial x}\delta n-\frac{\partial \delta u}{\partial x} n \right)\Bigr|_{x=0,l}=\frac{U_\mathrm{s,n}}{D_\mathrm{n}} \\
\frac{j_\mathrm{p}}{-D_\mathrm{p}} \Bigr|_{x=0,l}=\left(\frac{\partial \delta p}{\partial x}+\frac{\partial u_0}{\partial x}\delta p+\frac{\partial \delta u}{\partial x} p \right)\Bigr|_{x=0,l}=\frac{U_\mathrm{s,p}}{D_\mathrm{p}}
\end{gathered}
\end{equation}
These equations determine the gradients of $\delta n$ and $\delta p$, or equivalently, the diffusion of excess electrons and holes, at the boundaries. The potential gradients, $\partial u_0/\partial x|_{x=0,l}$ and $\partial \delta u/\partial x|_{x=0,l}$ are non-zero only in the presence of surface charges or equivalently, surface states. In particular, the excess potential gradient is determined from the general relation
\begin{equation}
\begin{gathered}
\label{eqn_a6}
\frac{\partial \delta u}{\partial x}\Bigr|_{x=a}=\frac{\partial \delta u}{\partial x}\Bigr|_{x=0}-\beta \frac{e}{\epsilon_0 \epsilon} \int_0^{a}{dx'}\left[δp(x',t)-δn(x',t)\right]
\end{gathered}
\end{equation}
for $0\leq a\leq l$. Then the global charge neutrality condition is stated as $\partial \delta u/\partial x|_{x=l}=0$. In the absence of surface charges at $x=0$, we have $\partial \delta u/\partial x|_{x=0}=\partial \delta u/\partial x|_{x=l}=0$. Also, $U_\mathrm{s,n}=R_\mathrm{s,n}-G_\mathrm{s,n},\ U_\mathrm{s,n}=R_\mathrm{s,n}-G_\mathrm{s,n}$ are the net recombination and generation rates of charge carriers occurring at the surface due to certain surface states, if there exist any. We discuss the meaning of these terms thoroughly in section \ref{appendix_surf}. Throughout the report, surface charges are assumed to exist only at $x=0$. 

\section{The uniform bulk}
\label{appendix_bulk}
Assuming no surface effects, the following conditions hold. 
\begin{equation}
\begin{gathered}
\label{eqn_a7}
u_0=\frac{\partial u_0}{\partial x} \rightarrow n_0=n_\mathrm{b},\ p_0=p_\mathrm{b} \\ 
\frac{\partial u_0}{\partial x} \Bigr|_{x=0}=0, \ U_\mathrm{s,n}=U_\mathrm{s,p}=0
\end{gathered}
\end{equation}

We define the bulk recombination rate suitable for the low excitation regime, $R_\mathrm{b}=B_\mathrm{eff} (np-n_\mathrm{i}^2 )=B_\mathrm{eff} (n_\mathrm{b} \delta p+p_\mathrm{b} \delta n+\delta n \delta p)$, where $B_\mathrm{eff}=(n_\mathrm{i} \tau_\mathrm{eff} )^{-1}$ is the recombination coefficient for the semiconductor in its intrinsic state. $\tau_\mathrm{eff}$ is interpreted as the effective intrinsic lifetime of charge carriers (in the sense that it is the net effect of the band-to-band, Auger, and Shockley-Read-Hall type recombination processes) and can be treated as a constant for low excitations. Though not completely accurate, this is a good approximation for cases where surface effects are absent, and it can provide sufficient information about how bulk properties of the semiconductor are modified under different effective recombination rates without having to resort to numerical evaluation. The bulk (photo)generation process is assumed to be of Beer-Lambert type, $G_\mathrm{b}=N_0 \alpha_\mathrm{b} \mathrm{exp}(-\alpha_\mathrm{b} x)$, with the incident photon flux $N_0$ and wavelength-dependent bulk absorption coefficient $\alpha_\mathrm{b}$.

In the low excitation regime, the homogeneous semiconductor equations ($G_\mathrm{b}=0$) for the uniform bulk are reduced to
\begin{equation}
\begin{aligned}
\label{eqn_a8}
\frac{1}{D_\mathrm{n}} \frac{\partial \delta n}{\partial t}&=\frac{\partial^2 \delta n}{\partial x^2}-\frac{\partial \delta u}{\partial x} \frac{\partial \delta n}{\partial x}-\frac{\partial^2 \delta u}{\partial x^2} \delta n -n_\mathrm{b}\frac{\partial^2 \delta u}{\partial x^2}-\frac{R_\mathrm{b}}{D_\mathrm{n}} \\
&\approx \frac{\partial^2 \delta n}{\partial x^2}-n_\mathrm{b}\frac{\partial^2 \delta u}{\partial x^2}-\frac{R_\mathrm{b}}{D_\mathrm{n}} \\
&=\frac{\partial^2 \delta n}{\partial x^2}-\frac{\delta n}{S_\mathrm{n}^2}+\frac{\delta p}{K_\mathrm{n}^2}\\
\frac{1}{D_\mathrm{p}} \frac{\partial \delta p}{\partial t}&=\frac{\partial^2 \delta p}{\partial x^2}+\frac{\partial \delta u}{\partial x} \frac{\partial \delta p}{\partial x}+\frac{\partial^2 \delta u}{\partial x^2} \delta p + p_\mathrm{b}\frac{\partial^2 \delta u}{\partial x^2}-\frac{R_\mathrm{b}}{D_\mathrm{p}} \\
&\approx \frac{\partial^2 \delta p}{\partial x^2}+p_\mathrm{b}\frac{\partial^2 \delta u}{\partial x^2}-\frac{R_\mathrm{b}}{D_\mathrm{p}} \\
&=\frac{\partial^2 \delta p}{\partial x^2}-\frac{\delta p}{S_\mathrm{p}^2}+\frac{\delta n}{K_\mathrm{p}^2}
\end{aligned}
\end{equation}
with the new length parameters introduced in the final equations defined in Table~\ref{tbl_sem}. 

\begin{table}[h]
\caption{Definition of the length parameters in the low excitation regime for different types of semiconductors.}
\begin{tabular}{ccccc}
\label{tbl_sem}
Parameter & \begin{tabular}[c]{@{}c@{}}General\end{tabular} & \begin{tabular}[c]{@{}c@{}}Intrinsic\end{tabular} & \begin{tabular}[c]{@{}c@{}}p-type\end{tabular} & \begin{tabular}[c]{@{}c@{}}n-type\end{tabular} \\ \hline
$\frac{1}{S_\mathrm{n}^2}$    & $\frac{1}{\lambda_\mathrm{D_\mathrm{n}}^2}+\frac{1}{l_\mathrm{n}'^{2}}$                                                             & $\frac{1}{\lambda_\mathrm{D_\mathrm{n}}^2}+\frac{1}{l_\mathrm{n}^{2}}$                                                      & $\frac{1}{l_\mathrm{n}'^{2}}$                                                   & $\frac{1}{\lambda_\mathrm{D_\mathrm{n}}^2}$                                                   \\ \hline
$\frac{1}{K_\mathrm{n}^2}$    & $\frac{1}{\lambda_\mathrm{D_\mathrm{n}}^2}-\frac{1}{\chi_n l_\mathrm{n}^{2}}$                                                            & $\frac{1}{\lambda_\mathrm{D_\mathrm{n}}^2}-\frac{1}{l_\mathrm{n}^{2}}$                                                      & 0                                                   & $\frac{1}{\lambda_\mathrm{D_\mathrm{n}}^2}-\frac{1}{\chi_n l_\mathrm{n}^{2}}$                                                   \\ \hline
$\frac{1}{S_\mathrm{p}^2}$    & $\frac{1}{\lambda_\mathrm{D_\mathrm{p}}^2}+\frac{1}{l_\mathrm{p}'^{2}}$                                                            & $\frac{1}{\lambda_\mathrm{D_\mathrm{p}}^2}+\frac{1}{l_\mathrm{p}^{2}}$                                                     &  $\frac{1}{\lambda_\mathrm{D_\mathrm{p}}^2}$                                                 & $\frac{1}{l_\mathrm{p}'^{2}}$                                                  \\ \hline
$\frac{1}{K_\mathrm{p}^2}$    & $\frac{1}{\lambda_\mathrm{D_\mathrm{p}}^2}-\frac{1}{\chi_p l_\mathrm{p}^{2}}$                                                           & $\frac{1}{\lambda_\mathrm{D_\mathrm{p}}^2}-\frac{1}{l_\mathrm{p}^{2}}$                                                     & $\frac{1}{\lambda_\mathrm{D_\mathrm{p}}^2}-\frac{1}{\chi_p l_\mathrm{p}^{2}}$                                                  & 0                                                 
\end{tabular}
\end{table}

The semiconductor type is determined according to the following relation: intrinsic -- $\mathrm{O}(\delta n,\ \delta p)\ll n_\mathrm{b}=p_\mathrm{b}=n_\mathrm{i}$, p (n)-type -- $n_\mathrm{b}\ (p_\mathrm{b})\ll\mathrm{O}(\delta n,\ \delta p)\ll p_\mathrm{b}\ (n_\mathrm{b})$ where the big-O notation denotes the order of magnitude of the excess carrier densities. The definition for the intrinsic diffusion lengths is $l_\mathrm{n}=(D_\mathrm{n} \tau_\mathrm{eff})^{1/2},l_\mathrm{p}=(D_\mathrm{p} \tau_\mathrm{eff})^{1/2}$. The extrinsic diffusion lengths $l_\mathrm{n}',l_\mathrm{p}'$ are defined as 
\begin{equation}
\label{eqn_a9}
l_\mathrm{n}'=(D_\mathrm{n} (\chi_\mathrm{n} \tau_\mathrm{eff}))^{1/2},\ l_\mathrm{p}'= (D_\mathrm{p} (\chi_\mathrm{p} \tau_\mathrm{eff}))^{1/2}
\end{equation}
where $\chi_\mathrm{n}=n_\mathrm{i}/n_\mathrm{b},\ \chi_\mathrm{p}=n_\mathrm{i}/p_\mathrm{b}$ may be interpreted as weight factors that modify the intrinsic diffusion lengths to their extrinsic values in doped cases. Note that the Poisson equation has not been removed, but rather absorbed into the continuity equations. Therefore, the equations fully account for charge distribution effects. 

The solutions of Eq.~(\ref{eqn_a8}) can be solved using separation of variables with respect to space and time. Using the ansatz, $\delta n(x,t)=\delta n_x \delta n_t,\ \delta p(x,t)=\delta p_x \delta p_t$, we get
\begin{equation}
\begin{aligned}
\label{eqn_a10}
\frac{1}{D_\mathrm{n}} \frac{1}{\delta n_t} \frac{\partial \delta n_t}{\partial t}=\frac{1}{\delta n_x} \frac{\partial^2 \delta n_x}{\partial x^2}-\frac{1}{S_\mathrm{n}^2}+\frac{1}{K_\mathrm{n}^2}\frac{\delta p_x}{\delta n_x}\frac{\delta p_t}{\delta n_t}\\
\frac{1}{D_\mathrm{p}} \frac{1}{\delta p_t} \frac{\partial \delta p_t}{\partial t}=\frac{1}{\delta p_x} \frac{\partial^2 \delta p_x}{\partial x^2}-\frac{1}{S_\mathrm{p}^2}+\frac{1}{K_\mathrm{p}^2}\frac{\delta n_x}{\delta p_x}\frac{\delta n_t}{\delta p_t}
\end{aligned}
\end{equation}
Now, we substitute $δn_t=δp_t=\mathrm{exp}(-\gamma t)$ into the above equation. Then, $\delta p_t/\delta n_t=\delta n_t/\delta p_t=1$. Defining $E_\mathrm{n},E_\mathrm{p}$ as the constants associated to the separated variables, we obtain a set of equations
\begin{equation}
\begin{gathered}
\label{eqn_a11}
\frac{\partial \delta n_t}{\partial t}=-D_\mathrm{n} E_\mathrm{n} \delta n_t \\
\frac{\partial \delta p_t}{\partial t}=-D_\mathrm{p} E_\mathrm{p} \delta p_t \\
\frac{\partial^2 \delta n_x}{\partial x^2}-\left(\frac{1}{S_\mathrm{n}^2} - E_\mathrm{n}\right)\delta n_x=-\frac{\delta p_x}{K_\mathrm{n}^2} \\
\frac{\partial^2 \delta p_x}{\partial x^2}-\left(\frac{1}{S_\mathrm{p}^2} - E_\mathrm{p}\right)\delta p_x=-\frac{\delta n_x}{K_\mathrm{p}^2} \\
E_\mathrm{n}=\frac{\gamma}{D_\mathrm{n}}, \ E_\mathrm{p}=\frac{\gamma}{D_\mathrm{p}}
\end{gathered}
\end{equation}
Let us consider two limiting cases where we can develop intuition about the general solutions that are to be derived shortly. 

\subsection{Temporally stationary case}
\label{appendix_bulk_1}
The stationary spatial density of charge carriers may be obtained under the condition, $\partial \delta n_t/\partial t=\partial \delta p_t/\partial t=0 \Leftrightarrow \gamma=0$. Recovering $G_\mathrm{b}=N_0 \alpha_\mathrm{b} \mathrm{exp}(-\alpha_\mathrm{b} x)$ in the right-hand side of the equations, we obtain
\begin{equation}
\begin{aligned}
\label{eqn_a12}
\frac{\partial^2 \delta n_x}{\partial x^2}-\frac{\delta n_x}{S_\mathrm{n}^2}=-\frac{\delta p_x}{K_\mathrm{n}^2} + \frac{G_\mathrm{b}}{D_\mathrm{n}}\\
\frac{\partial^2 \delta p_x}{\partial x^2}-\frac{\delta p_x}{S_\mathrm{p}^2}=-\frac{\delta n_x}{K_\mathrm{p}^2} + \frac{G_\mathrm{b}}{D_\mathrm{p}}
\end{aligned}
\end{equation}
The homogeneous solutions are found with the ansatz, $\delta n_{x,h}=\delta n_x (0) \mathrm{exp}(-x/r),\ \delta p_{x,h}=\delta p_x (0) \mathrm{exp}(-x/r)$, while the particular solutions can be calculated with the ansatz, $\delta n_{x,p}=C_\mathrm{n} \mathrm{exp}(-\alpha_\mathrm{b} x),\ \delta p_{x,p}=C_\mathrm{p} \mathrm{exp}(-\alpha_\mathrm{b} x)$. The boundary conditions used to determine the coefficients in the homogeneous solution are
\begin{equation}
\begin{aligned}
\label{eqn_a13}
\frac{\partial \delta n}{\partial x}\Bigr|_{x=0}=\frac{\partial \delta p}{\partial x}\Bigr|_{x=0}=\frac{\partial \delta n}{\partial x}\Bigr|_{x=l}=\frac{\partial \delta p}{\partial x}\Bigr|_{x=l}=0.
\end{aligned}
\end{equation}
where global charge neutrality $\partial \delta u/\partial x|_{x=0}=\partial \delta u/\partial x|_{x=l}=0$ is implicit in the above expression. Through some algebra, the total solution is obtained as
\begin{widetext}
%\begin{equation}
\begin{align}
\label{eqn_a14} 
\begin{bmatrix}
\delta n_x \\
\delta p_x
\end{bmatrix}
=\begin{bmatrix}
\delta n_x \\
\delta p_x
\end{bmatrix}_{h}
+\begin{bmatrix}
\delta n_x \\
\delta p_x
\end{bmatrix}_{p}
=\begin{bmatrix}
v_1 & u_1 \\
v_2 & u_2
\end{bmatrix}
\begin{bmatrix}
A_{+}\mathrm{cosh}\left(\frac{x}{r_{+}}\right) + B_{+}\mathrm{sinh}\left(\frac{x}{r_{+}}\right) \\
A_{-}\mathrm{cosh}\left(\frac{x}{r_{-}}\right) + B_{-}\mathrm{sinh}\left(\frac{x}{r_{-}}\right)
\end{bmatrix}
+ \begin{bmatrix}
C_\mathrm{n} \mathrm{exp}(-\alpha_\mathrm{b} x) \\
C_\mathrm{p} \mathrm{exp}(-\alpha_\mathrm{b} x)
\end{bmatrix}
\end{align}
\begin{gather*}
A_{+}=\frac{N_0}{W}(u_1 C_\mathrm{p} - u_2 C_\mathrm{n})\frac{r_{+}\alpha_\mathrm{b}}{\mathrm{sinh}\left(\frac{l}{r_+}\right)}\left[\mathrm{cosh}\left(\frac{l}{r_+}\right) - e^{-\alpha_\mathrm{b} l}\right], \ B_{+}=-\frac{N_0}{W} (u_1 C_\mathrm{p} - u_2 C_\mathrm{n})r_{+}\alpha_\mathrm{b} \\
A_{-}=-\frac{N_0}{W}(v_1 C_\mathrm{p} - v_2 C_\mathrm{n})\frac{r_{-}\alpha_\mathrm{b}}{\mathrm{sinh}\left(\frac{l}{r_{-}}\right)}\left[\mathrm{cosh}\left(\frac{l}{r_{-}}\right) - e^{-\alpha_\mathrm{b} l}\right], \ B_{-}=\frac{N_0}{W} (v_1 C_\mathrm{p} - v_2 C_\mathrm{n})r_{-}\alpha_\mathrm{b} \\
C_\mathrm{n} = \frac{Y}{D_\mathrm{n}D_\mathrm{p}} \left[\frac{1}{K_n^2}D_\mathrm{n} + \left(\frac{1}{S_\mathrm{p}^2}  - \alpha_\mathrm{b}^2\right)D_\mathrm{p}\right], \ C_\mathrm{p} = \frac{Y}{D_\mathrm{n}D_\mathrm{p}}\left[\left(\frac{1}{S_\mathrm{n}^2}  - \alpha_\mathrm{b}^2\right)D_\mathrm{n} + \frac{1}{K_p^2}D_\mathrm{p}\right] \\
W=v_1 u_2 - v_2 u_1, \ Y=N_0 \alpha_\mathrm{b} \left[\left(\frac{1}{S_\mathrm{n}^2}  - \alpha_\mathrm{b}^2\right)\left(\frac{1}{S_\mathrm{p}^2}  - \alpha_\mathrm{b}^2 \right)-\frac{1}{K_n^{2} K_p^{2}}\right]^{-1}
\end{gather*}
%\end{equation}
\end{widetext}
where $r_{\pm}=1/\sqrt{\xi_{\pm}}$ are the spatial mode eigenvalues with 
\begin{equation}
\label{eqn_a15} 
\xi_{\pm}=\frac{1}{2}\left[\left(\frac{1}{S_\mathrm{n}^{2}} + \frac{1}{S_\mathrm{p}^{2}}\right) \pm \sqrt{\left(\frac{1}{S_\mathrm{n}^{2}} - \frac{1}{S_\mathrm{p}^2}\right)^{2} + \frac{4}{K_\mathrm{n}^{2}K_\mathrm{p}^{2}}}\right]
\end{equation}
and $v_+=\left[v_1 \ \ v_2\right]^{\mathrm{T}}, u_-=\left[u_1 \ \ u_2\right]^{\mathrm{T}}$ are the corresponding eigenvectors. These spatial modes are inherent bulk properties of the semiconductor with distinct physical significance. Borrowing terminologies from the reference \cite{krcmar_exact_2002}, where such spatial densities have been studied in the context of the Dember effect \cite{kronik_surface_1999}, $r_+$ corresponds to the Debye-screening mode and $r_-$ the diffusion-recombination mode. Note that the total charge density $\delta p_x - \delta n_x$ is non-zero, which cannot be derived from local charge neutrality. This implies that even in the absence of externally applied fields, illumination of light can charge a semiconductor. In general, this bulk charging increases with larger absorption coefficients $\alpha_\mathrm{b}$ and varies as a function of the material properties such as the intrinsic carrier concentration, carrier mobility, and doping concentration.

\subsection{Spatially flat case}
\label{appendix_bulk_2}
The temporal evolution of a spatially flat density can be solved under the condition, $\partial^2 \delta n_x/\partial x^2=\partial^2 \delta p_x/\partial x^2=0$. We consider the case where $G_\mathrm{b}=0$ with initially finite carrier densities, $\delta n_t,\ \delta p_t$. The solutions can be solved for in a similar fashion as the temporally stationary case, which are obtained as
\begin{equation}
\label{eqn_a16}
\begin{bmatrix}
\delta n_t \\
\delta p_t
\end{bmatrix}
=\begin{bmatrix}
\eta_1 & \sigma_1\\
\eta_2 &  \sigma_2
\end{bmatrix}
\begin{bmatrix}
T_{+}e^{-\gamma_+ t} \\
T_{-}e^{-\gamma_- t}
\end{bmatrix}
\end{equation}
where $\gamma_{\pm}$ are the temporal mode eigenvalues,
\begin{equation}
\label{eqn_a17}
\begin{gathered}
\gamma_{\pm} = \frac{1}{2}\left[\left(\frac{D_\mathrm{n}}{S_\mathrm{n}^{2}} + \frac{D_\mathrm{p}}{S_\mathrm{p}^{2}}\right) \pm \sqrt{\left(\frac{D_\mathrm{n}}{S_\mathrm{n}^{2}} - \frac{D_\mathrm{p}}{S_\mathrm{p}^2}\right)^{2} + \frac{4D_\mathrm{n}D_\mathrm{p}}{K_\mathrm{n}^{2}K_\mathrm{p}^{2}}}\right]
\end{gathered}
\end{equation}
and $\eta_+=\left[\eta_1 \ \ \eta_2\right]^{\mathrm{T}}, \sigma_-=\left[\sigma_1 \ \ \sigma_2\right]^{\mathrm{T}}$ are the corresponding eigenvectors. The total charge density can then be expressed as
\begin{equation}
\begin{gathered}
\delta p_t - \delta n_t = (\eta_2 - \eta_1)T_{+}e^{-\gamma_{+}t} + (\sigma_2 - \sigma_1)T_{-}e^{-\gamma_{-}t}
\end{gathered}
\end{equation}
Again, such an expression cannot be derived under the local charge neutrality condition. As in the temporally stationary case, the time constants associated with the eigenvalues $\tau_{\pm}=1/\gamma_{\pm}$ have distinct physical meanings, $\tau_{+}$ being the dielectric relaxation time, and $\tau_{-}$ the carrier lifetime \cite{roosbroeck_space_1961}. This is because the total charge density $\delta p_t-\delta n_t$ relaxes to zero in a characteristic time $\tau_{+}$, whereas the individual charge carrier densities $\delta n_t,\ \delta p_t$ diminish through diffusion and recombination over the characteristic time $\tau_{-}$. It can be interpreted that local charge neutrality ($\delta n_t=\delta p_t$) is achieved in time $\tau_{+}$, and that the system returns to thermal equilibrium ($\delta n_t=\delta p_t=0$) in time $\tau_{-}$. Given the relation Eq.~(\ref{eqn_a6}), the electric field reaches a constant value after $\tau_{+}$. In the absence of surface charges, the field is exactly zero, which means that the system completely neutralizes in the dielectric relaxation time. This is not true in the presence of surface charges. Non-uniform carrier trapping sites in the bulk can also complicate the dynamics. The additional charge equilibration processes introduced by material inhomogeneities or discontinuities of the material can modify the neutralization time from that of a uniform bulk.

\begin{figure}[h]
\centering
\includegraphics[width=0.8\columnwidth]{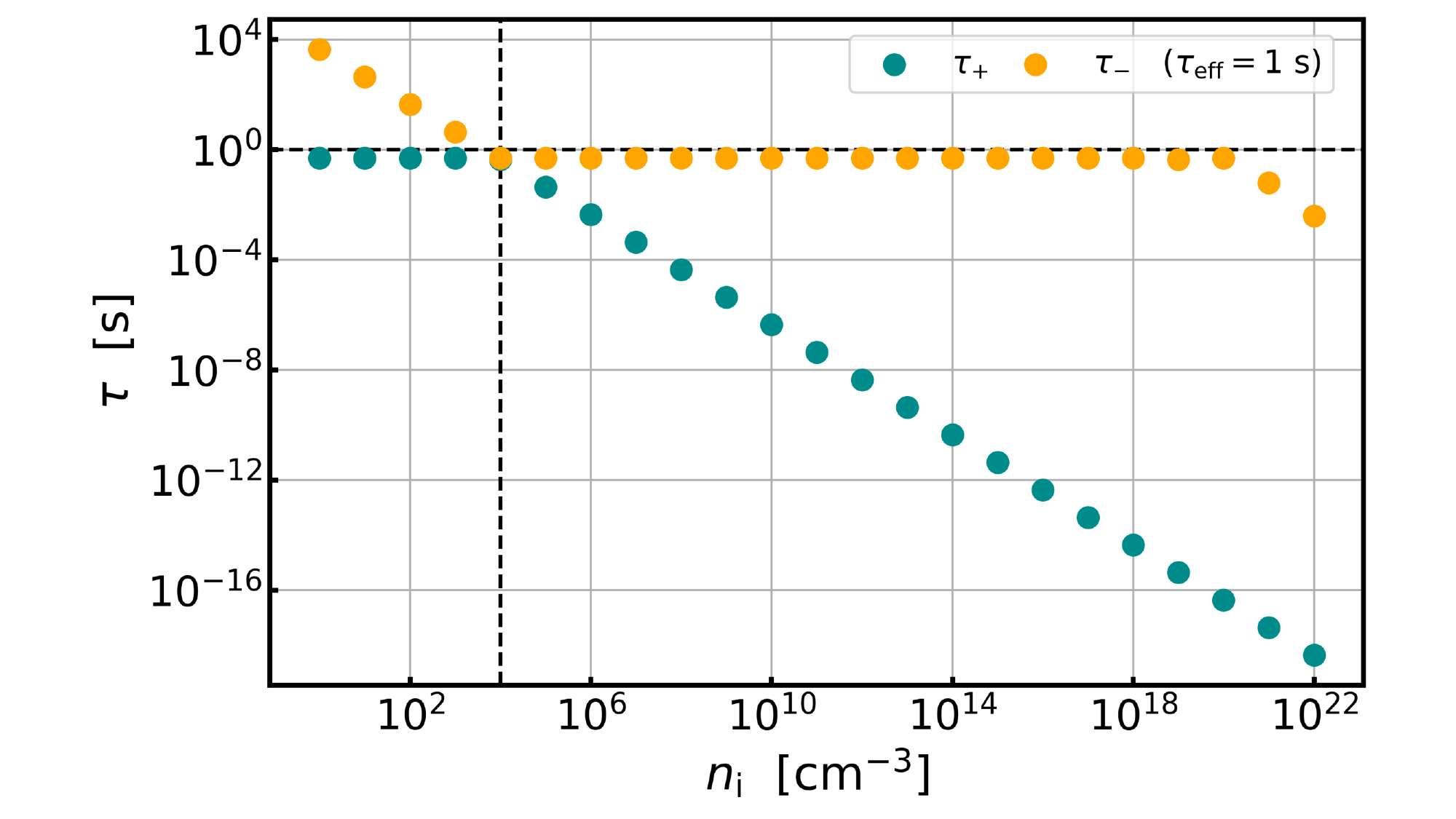}
\caption{Plot of $\tau_+$ and $\tau_{-}$ for varying $n_\mathrm{i}$. The horizontal dashed line is set by $\tau_\mathrm{eff} =$ 1 s, while the vertical dashed line indicates the near-degenerate point of $\tau_+$ and $\tau_{-}$.}
\label{fig_relaxation}
\end{figure}

Fig.~\ref{fig_relaxation} shows a plot of $\tau_+$ and $\tau_{-}$ defined in Eq.~(\ref{eqn_a17}) for a hypothetical material as a function of the intrinsic carrier concentration $n_\mathrm{i}$, with the effective carrier lifetime and carrier mobility values set to $\tau_\mathrm{eff}=$ 1 s and  $\mu_\mathrm{n} \ (\mu_\mathrm{p})=1000 \ (300)$ cm$^{2} \cdot$ V$^{-1}\cdot$ s$^{-1}$, respectively. The left and right regions of the plot correspond to the insulator (small $n_\mathrm{i}$) and conductor (large $n_\mathrm{i}$), while the middle region is indicative of the semiconductor (intermediate $n_\mathrm{i}$). The maximum value of $\tau_+$ is set by $\tau_\mathrm{eff}$, indicated as the horizontal dashed line. Values of $\mu_\mathrm{n}$ and $\mu_\mathrm{p}$ determine the location of the crossing point (or degenerate point) between $\tau_+$ and $\tau_{-}$, shifting the location of the vertical line.

Note that the dielectric relaxation of an unpaired charge density $\rho_\mathrm{u}$ in conductors and insulators is described by the equation

\begin{equation}
\begin{gathered}
\label{eqn_dielectric}
\frac{\partial \rho_\mathrm{u}}{\partial t} + \frac{\rho_\mathrm{u}}{\epsilon_0\epsilon/\sigma} = \frac{\partial \phi}{\partial x} \left(\frac{\partial \sigma}{\partial x} - \frac{\sigma}{\epsilon}\frac{\partial \epsilon}{\partial x}\right)
\end{gathered}
\end{equation}
which is basically the continuity equation for inhomogeneous ohmic materials \cite{haus_em_1989}. When the dielectric constant $\epsilon$ and conductivity $\sigma$ is homogeneous throughout the material, the right-hand side of the equation vanishes and the dielectric relaxation time is obtained as $\tau_{+}=\epsilon_0\epsilon/\sigma$. In these systems, there are no dynamics of mobile carriers, in the sense that an initial density relaxes to the boundary without generating a net charge density beyond the initially occupied volume. \cite{haus_em_1989}.

\subsection{General spatiotemporal solutions}
\label{appendix_bulk_3}
Solving the coupled homogeneous equations Eq.~(\ref{eqn_a11}) for both space and time, it can be shown that the spatial and temporal modes, 1) the Debye-screening mode ($r_+$) and the dielectric relaxation time ($\tau_+$), and 2) the diffusion-recombination mode ($r_-$) and the carrier lifetime ($\tau_-$), are directly coupled. This presents a consistent framework for the bulk response of a semiconductor in non-equilibrium. The general solutions are obtained as
\begin{widetext}
%\begin{equation}
\begin{align}
\label{eqn_a18} 
\begin{bmatrix}
\delta n(x,t) \\
\delta p(x,t)
\end{bmatrix}
=\mathlarger{\mathlarger{\sum}}_{m_{+}, m_{-}}
\begin{bmatrix}
v_{1,m_+} & u_{1,m_-} \\
v_{2,m_+} & u_{2,m_-}
\end{bmatrix}
\begin{bmatrix}
e^{-\gamma_{+}t}\left\{A_{m_+}\mathrm{cosh}\left(\frac{m_+ \pi}{l}\right) + B_{m_+}\mathrm{sinh}\left(\frac{m_+ \pi}{l}\right)\right\} \\
e^{-\gamma_{-}t}\left\{A_{m_-}\mathrm{cosh}\left(\frac{m_- \pi}{l}\right) + B_{m_-}\mathrm{sinh}\left(\frac{m_- \pi}{l}\right) \right\}
\end{bmatrix}
\end{align}
\begin{gather*}
\gamma_{\pm}=\left\{\begin{matrix}
\gamma_+ = \frac{1}{2}\left\{\left[\left(\frac{D_\mathrm{n}}{S_\mathrm{n}^2} + \frac{D_\mathrm{p}}{S_\mathrm{p}^2} \right) + \left(\frac{m_+ \pi}{l}\right)^{2} (D_\mathrm{n} + D_\mathrm{p})\right] + \sqrt{\left[\left(\frac{D_\mathrm{n}}{S_\mathrm{n}^2} - \frac{D_\mathrm{p}}{S_\mathrm{p}^2} \right) + \left(\frac{m_+ \pi}{l}\right)^{2} (D_\mathrm{n} - D_\mathrm{p})\right]^{2} + \frac{4D_\mathrm{n}D_\mathrm{p}}{K_\mathrm{n}^{2}K_\mathrm{p}^{2}}} \right\} \\
\gamma_- = \frac{1}{2}\left\{\left[\left(\frac{D_\mathrm{n}}{S_\mathrm{n}^2} + \frac{D_\mathrm{p}}{S_\mathrm{p}^2} \right) + \left(\frac{m_- \pi}{l}\right)^{2} (D_\mathrm{n} + D_\mathrm{p})\right] - \sqrt{\left[\left(\frac{D_\mathrm{n}}{S_\mathrm{n}^2} - \frac{D_\mathrm{p}}{S_\mathrm{p}^2} \right) + \left(\frac{m_- \pi}{l}\right)^{2} (D_\mathrm{n} - D_\mathrm{p})\right]^{2} + \frac{4D_\mathrm{n}D_\mathrm{p}}{K_\mathrm{n}^{2}K_\mathrm{p}^{2}}} \right\}
\end{matrix} \right.
\end{gather*}
%\end{equation}
\end{widetext}
where $m_{\pm}\in \mathbb{Z}$. This expression is a generalization of the limiting case solutions presented in the previous sections. The spatial modes are split into the Debye-screening ($m_+$) and diffusion-recombination ($m_-$) modes to which are associated the characteristic time constants $\tau_+ =1/\gamma_{m_+}$ (generalized dielectric relaxation time) and $\tau_- =1/\gamma_{m_-}$ (generalized carrier lifetime), respectively. In order to describe the most general charge carrier dynamics, including a generation process $G_\mathrm{b}$, we can use the Fourier series analysis using the homogeneous solutions $\left[\delta n(x,t) \ \ \delta p(x,t))\right]^T$ and determine the coefficients, $A_\mathrm{\pm},B_\mathrm{\pm}$.

\section{The presence of surface charges}
\label{appendix_surf}
Analytical solutions are not obtainable in the presence of surface charges because space charge quantities in thermal equilibrium are not constant, i.e., $u_0,\ \partial u_{0}/\partial x \neq 0$, hence rendering the semiconductor equations nonlinear even in the low excitation regime. Therefore, the semiconductor equations Eq.~(\ref{eqn_a4}) must be solved numerically.

Surface charges typically originate from surface states, and can largely be classified into two categories \cite{garrett_physical_1955, johnson_large-signal_1958}. The first is fixed surface charge, which is long-term fixed charge that remains stationary during the dynamics of excess charge carriers in non-equilibrium, commonly associated with the slow surface state. The second is charge that originates from the interface state (or fast surface state) which is basically a Shockley-Read-Hall type defect state within the bandgap of the semiconductor localized at the surface that can be exchanged with the bulk. We denote the charge densities associated with the fixed surface charge and interface state as $\Sigma_\mathrm{ss}$ and $\Sigma_\mathrm{fs}$, respectively. Since nonzero surface charge density gives rise to a potential gradient at the surface, boundary condition values (see equation Eq.~(\ref{eqn_a5})) that were nulled in the uniform bulk problem must be recovered. The potential gradient can be decomposed into the equilibrium and excess potential gradients, and then into the contributions from the fixed surface charge and interface state as
\begin{equation}
\label{eqn_a19}
\begin{gathered}
\frac{\partial u}{\partial x}\Bigr|_{x=0} = \frac{\partial u_0}{\partial x}\Bigr|_{x=0} + \frac{\partial \delta u}{\partial x}\Bigr|_{x=0} \\
\frac{\partial u_0}{\partial x}\Bigr|_{x=0} = \frac{\partial u_\mathrm{ss}}{\partial x}\Bigr|_{x=0} + \frac{\partial u_\mathrm{fs,0}}{\partial x}\Bigr|_{x=0} \\
\frac{\partial \delta u}{\partial x}\Bigr|_{x=0}= \frac{\partial \delta u_\mathrm{fs}}{\partial x}\Bigr|_{x=0}
\end{gathered}
\end{equation}
where we used $u_\mathrm{ss}=u_\mathrm{ss,0}$, and $\delta u_\mathrm{fs}=u_\mathrm{fs}-u_\mathrm{fs,0}$. A fixed surface charge of $\pm e\Sigma_\mathrm{ss}$ results in the equilibrium potential gradient
\begin{equation}
\label{eqn_a20}
\frac{\partial u_\mathrm{ss}}{\partial x}\Bigr|_{x=0} = \mp\beta\frac{e}{\epsilon \epsilon_0}\Sigma_\mathrm{ss}
\end{equation}
On the other hand, the interface state is characterized by numerous parameters. Let us consider two types of discrete interface states: an acceptor-type and a donor-type. The acceptor-type is negative (neutral) when occupied by an electron (a hole), whereas the donor-type is neutral (positive) when occupied by an electron (a hole). The potential gradient is given as
\begin{equation}
\label{eqn_a21}
\frac{\partial u_\mathrm{fs}}{\partial x}\Bigr|_{x=0} = \pm\beta\frac{e}{\epsilon \epsilon_0}\Sigma_\mathrm{fs}\times \left \{ \begin{matrix}f_\mathrm{s} & & \textrm{acceptor-type} \\ 1-f_\mathrm{s} & &  \textrm{donor-type} \end{matrix} \right.
\end{equation}
where $f_\mathrm{fs}$ is the electron occupation probability of the interface state. We focus on two processes that may occur through these states, 1) surface recombination and 2) surface absorption (also known as photoionization), in the presence of which the general rate equation associated with $f_\mathrm{s}$ is given as \cite{shockley_recomb_1952, hall_recomb_1952, hsieh_recomb_1989}
\begin{equation}
\label{eqn_a22}
\begin{gathered}
\frac{\partial f_\mathrm{s}}{\partial t}=U_\mathrm{s,n}-U_\mathrm{s,p}, \ U_\mathrm{s,n}=R_\mathrm{s,n}-G_\mathrm{s,n},\ U_\mathrm{s,p}=R_\mathrm{s,p}-G_\mathrm{s,p} \\
R_\mathrm{s,n}=s_\mathrm{n0} n (1-f_\mathrm{s} ), \ G_\mathrm{s,n}=(s_\mathrm{n0} n_1 + n^\mathrm{o} ) f_\mathrm{s} \\
R_\mathrm{s,p}=s_\mathrm{p0} p f_\mathrm{s},\ G_\mathrm{s,p}=(s_\mathrm{p0} p_1 + p^\mathrm{o})(1-f_\mathrm{s})
\end{gathered}
\end{equation}
Recall $U_\mathrm{s,n},\ U_\mathrm{s,p}$ from section \ref{appendix_sem_eq}. 
The trap parameters $ s_\mathrm{n0}=\sigma_\mathrm{n}^\mathrm{c} \Sigma_\mathrm{fs} v_\mathrm{n},\ s_\mathrm{p0}=\sigma_\mathrm{p}^\mathrm{c} \Sigma_\mathrm{fs} v_\mathrm{p}$ are the electron and hole surface recombination velocities with dimensions $[s_\mathrm{n0}]=[s_\mathrm{p0}]=$ cm$\cdot$s$^{-1}$. $\sigma_\mathrm{n}^\mathrm{c},\ \sigma_\mathrm{p}^\mathrm{c}$ and $v_\mathrm{n},\ v_\mathrm{p}$ are the capture cross sections and thermal velocities of electrons and holes whose dimensions are $[\sigma_\mathrm{n}^\mathrm{c}]=[\sigma_\mathrm{p}^\mathrm{c}]=$ cm$^{2}$ and $[v_\mathrm{n}]=[v_\mathrm{p}]=$ cm$\cdot$s$^{-1}$, respectively. The concentrations $n_{1}=n_\mathrm{i} \mathrm{exp}({-u_\mathrm{fs}}),\ p_{1}=n_\mathrm{i} \mathrm{exp}({u_\mathrm{fs}})$ are determined by the energy level of the defect $u_\mathrm{fs}=\beta \phi_\mathrm{fs}$ within the bandgap. Introducing the dimensionless surface absorption coefficients $\alpha_\mathrm{n}^\mathrm{o}=\sigma_\mathrm{n}^\mathrm{o} \Sigma_\mathrm{fs},\ \alpha_\mathrm{p}^\mathrm{o}=\sigma_\mathrm{p}^\mathrm{o} \Sigma_\mathrm{fs}$ where $\sigma_\mathrm{n}^\mathrm{o},\ \sigma_\mathrm{p}^\mathrm{o}$ are the optical cross sections for electrons and holes, we define the corresponding surface flux quantities, $n^\mathrm{o}=\alpha_\mathrm{n}^\mathrm{o} N_0,\ p^\mathrm{o}=\alpha_\mathrm{p}^\mathrm{o} N_0$ where $N_0$ is the incident photon flux used previously in the bulk generation process $G_\mathrm{b}$. The theory of optical cross sections is presented in the next section.
The terms $R_\mathrm{s,n},\ R_\mathrm{s,p}$ describe surface recombination, or capturing of free charge carriers from the bulk into the interface states, whereas $G_\mathrm{s,n},\ G_\mathrm{s,p}$ denote the release of captured charge carriers into the bulk. In particular, the first and second terms in $G_\mathrm{s,n},\ G_\mathrm{s,p}$ indicate thermal emission and optical generation rates, respectively, where the latter corresponds to surface absorption or photoionization \cite{hsieh_recomb_1989}. Here, we limit our analysis to steady state solutions, $\partial \delta n/\partial t=\partial \delta p/\partial t=\partial f_\mathrm{s}/∂t=0$, which results in a steady state value for the electron occupation probability $\bar{f_\mathrm{s}}$ and net charge carrier flow rate $U_\mathrm{s,n}=U_\mathrm{s,p}=U_\mathrm{s}$ as
\begin{equation}
\label{eqn_a23}
\begin{gathered}
\bar{f_\mathrm{s}}=\frac{(\frac{n}{s_\mathrm{p0}} +\frac{p_1^*}{s_\mathrm{n0}})}{\frac{1}{s_\mathrm{p0}}  (n+n_1^* )+\frac{1}{s_\mathrm{n0}} (p+p_1^* )}\Biggr|_{x=0} \\ 
U_\mathrm{s}=\frac{(np-n_1^* p_1^*)}{\frac{1}{s_\mathrm{p0}}  (n+n_1^* )+\frac{1}{s_\mathrm{n0}} (p+p_1^* )}\Biggr|_{x=0}
\end{gathered}
\end{equation}
where $n_1^*=n_1+n^\mathrm{o}/s_\mathrm{n0},\ p_1^*=p_1+p^\mathrm{o}/s_\mathrm{p0}$. In the presence of the interface states, then, the boundary value of the excess potential gradient is modified as
\begin{equation}
\label{eqn_a24}
\begin{gathered}
\frac{\partial \delta u}{\partial x}\Bigr|_{x=0}=\beta \frac{e}{\epsilon \epsilon_{0}} \Sigma_\mathrm{fs} \delta f_\mathrm{s}
\end{gathered}
\end{equation}
with $\delta f_\mathrm{s}=\bar{f_\mathrm{s}}-\bar{f_\mathrm{s}}_0$. Fixed surface charges do not contribute to this quantity since they are stationary and thus cancel out. Global charge neutrality $\partial \delta u/\partial x|_{x=l}=0$ must apply in order to balance charge transfer between the surface and bulk. This is naturally embedded in the relation
\begin{equation}
\label{eqn_a25}
\begin{gathered}
\frac{\partial \delta u}{\partial x}\Bigr|_{x=a}=\beta \frac{e}{\epsilon_0 \epsilon}\left\{\Sigma_\mathrm{fs} \delta f_\mathrm{s}-\int_0^{a}{dx'}\left[δp(x',t)-δn(x',t)\right]\right\}
\end{gathered}
\end{equation}
which is just equation Eq.~(\ref{eqn_a6}) expressed in terms of equation Eq.~(\ref{eqn_a24}). Equipped with the extended boundary conditions for $\partial u_0/\partial x|_{x=0}$ and $\partial \delta u/\partial x|_{x=0}$, the steady state solutions of the semiconductor equations can be readily obtained using numerical methods.

\section{Surface absorption and the optical cross section}
\label{appendix_abs}
Here, we briefly summarize the theoretical results presented in Ref.~\cite{ilaiwi_ocs_1990, tomak_ocs_1982}. The ground state wave function of the Hulth\'{e}n potential Eq.~(\ref{eqn_hulthen}) is 
\begin{equation}
\label{eqn_ground}
\braket{x|i}=\psi(x)=\left(\frac{4-\lambda^{2}}{4\pi\lambda^{2}a}\right)^{1/2} e^{-x/a} \frac{e^{x/2a} - e^{-x/2a}}{x}
\end{equation}
The photoionization cross section, in terms of the photon energy $\hbar \omega$, is obtained as
\begin{multline}
\label{eqn_ocs}
\sigma(\hbar\omega)=\left[\left(\frac{E_\mathrm{eff}}{E_0}\right)^{2}\frac{n(\hbar\omega)}{\epsilon}\right]\frac{16\pi \alpha}{3}\frac{\hbar \omega}{E_\mathrm{io}}\left(\frac{\hbar \omega }{E_\mathrm{io}} - 1\right)^{3/2}a^2 \\
\times c^{5/2}a^{5}\frac{4-\lambda^2}{\lambda^2} 
\Biggr\{\left[\left(1-\frac{\lambda}{2}\right)^{2}+ca^{2}\left(\frac{\hbar\omega}{E_\mathrm{io}}-1\right)\right]^{-2} \\
-\left[\left(1+\frac{\lambda}{2}\right)^{2}+ca^{2}\left(\frac{\hbar\omega}{E_\mathrm{io}}-1\right)\right]^{-2} \Biggr\}^{2}
\end{multline}
where $E_\mathrm{eff}/E_0$ is the effective field ratio, $n(\hbar \omega)$ the frequency-dependent refractive index, $\epsilon$ the dielectric constant of the material, $\alpha$ the fine structure constant, and $E_\mathrm{io}$ is the ionization energy between the defect level and the conduction band edge. We assume $E_\mathrm{eff}/E_0 \sim 2$ \cite{anderson_ocs_1975}, and replace $n(\hbar \omega)$ with an approximate average value of 4 for the experimental wavelengths \cite{pierce_optical_1972, aspnes_optical_1983}. The parameter $c$ is defined as $c=2m^{*}E_\mathrm{io}/\hbar^2$ where $m^{*}$ is the effective mass of the optically excited particle, which in our case, is the electron, $m_0$. In our calculations, we use $m^{*}=0.26m_0$ \cite{chenming_sem_2009}.

\section{The simulation geometry and estimation of the SPV}
\label{appendix_sim}
\begin{figure}[ht]
\centering
\includegraphics[width=1\columnwidth]{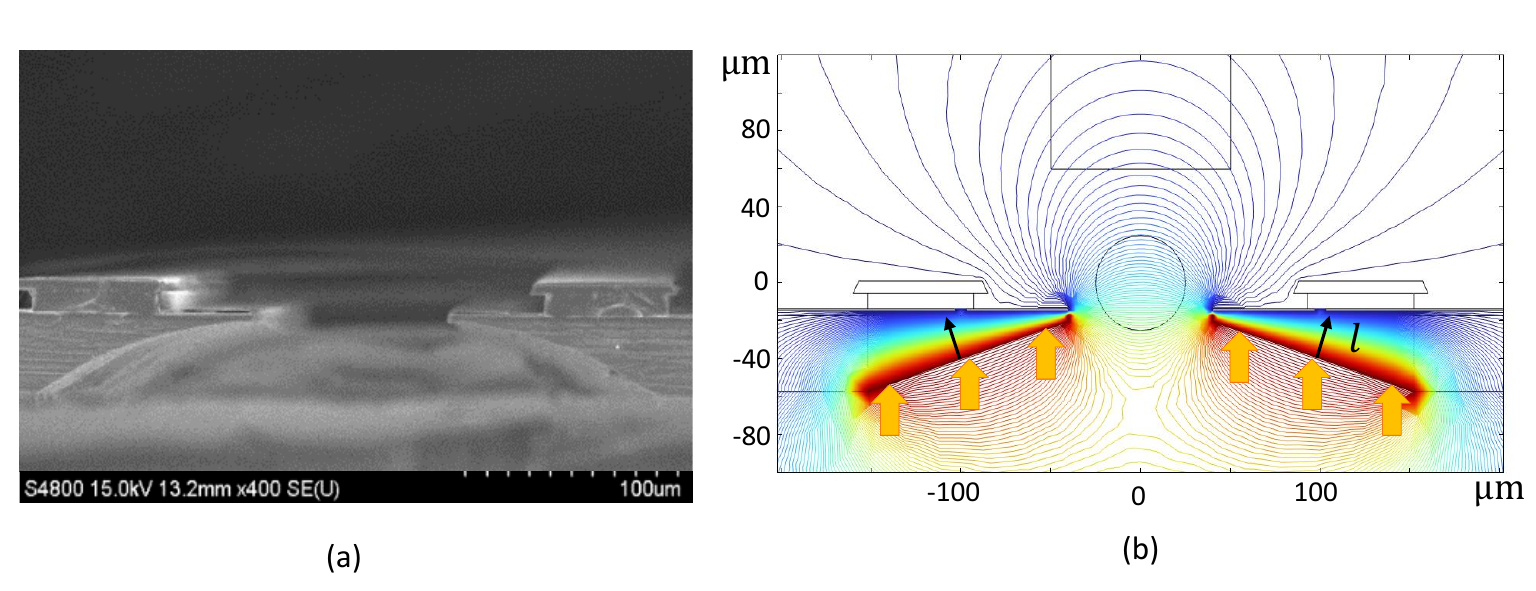}
\caption{SEM image and COMSOL geometry for numerical simulations. (a) An SEM image of the cross section of the ion trap near the trapping region. (b) COMSOL simulation geometry corresponding to (a) with equi-potential lines plotted when a voltage of +1 V is formed on the inclined surface of the silicon substrate (in red). Yellow arrows represent incident light and $l$ the semiconductor slab thickness.}
\label{fig_sim_appendix}
\end{figure}

In order to estimate the magnitude and sign of the stray field generated at the position of the trapped ion due to the SPV, an electrostatic analysis was performed using the COMSOL software by imposing voltages on the inclined surfaces of the exposed silicon substrate of the ion trap (see Fig.~\ref{fig_sim_appendix} (b)). The geometry for COMSOL simulations was extracted from the Scanning-electron-microscope (SEM) image shown in Fig.~\ref{fig_sim_appendix} (a). A voltage of +1 V at the silicon surface generated an electric field of +1055 V$\cdot$m $^{-1}$ at the position of the ion. On the other hand, +1 V applied to the inner dc electrode pairs produced an electric field of +2880 V$\cdot$m $^{-1}$ at the ion position. The experimental values of the SPV could be estimated systematically by multiplying this ratio, 2880/1055$\approx$2.73, to the absolute value of the compensation voltages.

\section{Alternative mechanism for the observed ion displacement and limitations of the model}
\label{appendix_lim}
Here we discuss an 1) alternative mechanism for the ion displacement and 2) the limitations of our photoconductive charging model.

\subsection{Alternative mechanism}
We consider an additional SPV mechanism that can produce a positive stray field at the position of the ion. Positive photovoltage can be induced by a purely bulk response of the semiconductor (see section \ref{appendix_bulk}). In particular, when the carrier mobilities satisfy $\mu_\mathrm{n}>\mu_\mathrm{p}$, which is true for silicon, the underlying charge density resembles that of a dipole with the positive side facing the surface of illumination (the Dember effect). The magnitude of the photovoltage, however, is small and decreases with increasing doping concentration due to enhanced screening (reduction in the Debye-screening length). In addition, the dielectric relaxation time, which is the characteristic time at which the semiconductor bulk neutralizes when light is turned off, is much shorter than the observed time scale at which the ion returns to its equilibrium position (1 to 100 \textmu s). For p-type silicon doped with a concentration of 10$^{15}$  cm$^{-3}$, the magnitude and relaxation time of the photovoltage are on the order of 10 mV and 10 ps, respectively.

Moreover, numerical simulations show that any mechanism dominated by bulk absorption predicts larger magnitudes of SPV at shorter wavelengths in accordance with the bulk absorption spectrum of silicon, failing to explain the spectral response of the observed SPV. Only in the presence of the proposed interface states can the magnitude, sign, and wavelength dependence of the SPV be comprehended consistently. Although this effect is small in our system, the photovoltage may still be problematic in others depending on doping concentration and the proximity of the trapped ion to the charged volume.

\subsection{Limitations of the photoconductive charging model}
The following list states some limitations of our photoconductive charging model.

1. Interface states have been assumed to occupy a single discrete energy level, while in more realistic systems, they would more likely form a distribution within the bandgap, in which case the observations would be more of a collective response. Although the former assumption allows for an effective explanation and a more tractable computation, future work may be devoted to studying more generalized surface conditions involving a distribution of interface states.

2. The theory of photoionization from bulk defects has been applied to surface defects. Although a complete theory for the optical excitation of surface defects is lacking \cite{kronik_surface_1999}, the optical cross section of a delta-function defect as a function of the distance from a surface has been studied in Ref.~\cite{chazalviel_abs_1987}, showing a tendency in the optical cross section spectrum to broaden, while its peak value is shifted to larger photon energies, as the defect becomes closer to the surface. The fitted values of $a$ and $\lambda$ in the Hulth\'{e}n potential may be slightly modified if such effects are included in the model. Since the overall spectral dependence of the SPV observed in our experiments are explained well by that derived for bulk defects, we suspect the optically responsive interface states to have originated from RIE-induced defects that penetrated deep enough into the substrate to have spectral properties resembling that of the bulk defect, but sufficiently localized at the surface (i.e., within a few atomic layers to several nm's, which is much smaller than the Debye-screening length and the absorption depth of any incident light) so that their effects are manifest as boundary conditions in the context of the semiconductor equations.

3. We have assumed a slab model, but this may not be able to fully describe the real exposed surface of the silicon substrate which is a more complicated three-dimensional structure. One justification for using the slab model was based on the numerical simulation results that were insensitive to a variation of the slab thickness l as long as it was much larger than the initial surface depletion layer ($\sim$1 \textmu m). However, even in this case, edge effects or diffusion and drift in spatial dimensions other than in the direction of incidence of light were neglected.

4. External field effects have been neglected from the model. This was justified by the experimental fact that the sign and magnitude of the SPV were independent of the changes in voltages applied to the dc electrodes in the vicinity of the exposed semiconductor surface.

\section{Simulation of the quantum dynamics}
\label{appendix_quantum_sim}
Here, the theory used for simulations of the Rabi oscillation and Bloch sphere trajectory is presented.

\subsection{Lindblad master equation}
\label{appendix_lindblad}
The trapped ion is a composite system, involving the qubit and oscillator degrees of freedom. Its density matrix can be expressed as $\rho(t)={\sum}_{m,n}\rho^{(m,n)}(t)\otimes\ket{m}\bra{n}$ where $\rho^{(m,n)}(t)$ is the qubit state corresponding to the subspace formed by the oscillator eigenstates $\ket{m}$ and $\ket{n}$. When the oscillator is coupled to a phonon bath via an amplitude damping channel described by the Lindblad operator, $L = \Gamma a $, the Lindblad master equation can be formulated as
\cite{gardiner_1989}

\begin{widetext}
\begin{equation}
\label{eqn_lindblad}
\begin{aligned}
d\rho^{(m,n)}=&-\frac{i}{\hbar}\left(\sum_{r}{H_\mathrm{sys}}_{m,r}\rho^{(r,n)}-\sum_{l}\rho^{(m,l)}{H_\mathrm{sys}}_{l,n}\right) 
\\&-\frac{\Gamma}{2}\left((2\bar{n}_\mathrm{T}+1)(m+n)+2\bar{n}_\mathrm{T}\right)\rho^{(m,n)}+\Gamma(\bar{n}_\mathrm{T}+1)\sqrt{(m+1)(n+1)}\rho^{(m+1,n+1)}+\Gamma\sqrt{mn}\rho^{(m-1,n-1)}
\end{aligned}
\end{equation}
\end{widetext}

where $H_\mathrm{sys}$ is the system Hamiltonian, $\Gamma$ is the heating rate, and $\bar{n}_\mathrm{T}$ is the mean phonon number of the phonon bath evaluated in terms of the oscillator states. We set the initial condition as $\rho(0)=\ket{0}\bra{0}\otimes{\sum}_{n}P_n(\bar{n}_0)\ket{n}\bra{n}$ where the qubit state is initialized to $\ket{0}$ and the oscillator state has a thermal distribution $P_n(\bar{n}_0)={\bar{n}_0}^n/\left(1+\bar{n}_0\right)^{n+1}$ about the mean phonon number $\bar{n}_0$. We propagate the state through time using the update rule, $\rho^{(m,n)}(t+dt)=\rho^{(m,n)}(t)+d\rho^{(m,n)}(t)$, and obtain the reduced density matrix describing the qubit state by taking the partial trace over the oscillator states, $\rho_\mathrm{qubit}(t)={\sum}_{n}\rho^{(n,n)}(t)$. 

\subsection{Position operator of the time-dependent oscillator}
\label{appendix_td_oscillator}
The theory of forced time-dependent oscillators presented in Refs.\cite{ji_1995, ji_1996} is applied to the linear Paul trap. The Hamiltonian for the oscillator degree of freedom of the trapped ion is given as
\begin{equation}
\label{eqn_td_oscillator}
\begin{gathered}
H(t)=H_{0}(t)+V(t) \\
H_{0}(t) = \frac{p^2}{2M} + \frac{1}{2}MW^{2}(t)x^2 \\
V(t) = -F(t)x
\end{gathered}
\end{equation}
where $H_{0}(t)$ describes the dynamically trapped ion, and $F(t)$ is an externally driven force. $M$ is the mass of the ion. In the subsequent derivations, the Planck constant is set to $\hbar=1$, but recovered in the final expressions. In this system, the position operator in the Heisenberg picture is obtained as
\begin{equation}
\label{eqn_apos}
x(t)=\sqrt{\frac{g_{-}(t)}{2\omega_\mathrm{I}}}\left(e^{i\omega(t)}a^{\dag}+e^{-i\omega(t)}a+2\mathrm{Re}(\alpha(t))\right)
\end{equation}
where $a^{\dag}(a)$ are the raising(lowering) operators of a reference oscillator defined at $t=0$, and $\omega_\mathrm{I}$ is an invariant of motion, defined as
\begin{equation}
\label{eqn_invariant}
\omega_\mathrm{I}=\sqrt{g_{+}(t)g_{-}(t)-g^{2}_{0}(t)}.
\end{equation}
The parameters $g_{\pm}(t),\ g_{0}(t)$ are determined from the coupled first-order differential equations
\begin{equation}
\label{eqn_g}
\begin{gathered}
\dot{g}_{-}=-\frac{2}{M}g_{0} \\
\dot{g}_{0}=-MW^{2}(t)g_{-} - \frac{g_{+}}{M} \\
\dot{g}_{+}=2MW^{2}(t)g_{0}.
\end{gathered}
\end{equation}
The time-dependent frequency $\omega(t)$ is obtained as
\begin{equation}
\label{eqn_omega}
\omega(t)=\mathlarger{\int}_{0}^{t}dt'\frac{\omega_\mathrm{I}}{Mg_{-}(t')}
\end{equation}
while the displacement $\alpha(t)$ is derived as
\begin{equation}
\label{eqn_displacement}
\alpha(t)=e^{-i\omega(t)}\alpha(0)+i\mathlarger{\mathlarger{\int}}_{0}^{t}dt'e^{-i\left(\omega(t)-\omega(t')\right)}\sqrt{\frac{g_{-}(t')}{2\omega_\mathrm{I}}}F(t').
\end{equation}
With the definition of $W^{2}(t)$ for the linear Paul trap
\begin{equation}
\label{eqn_W}
W^{2}(t)=\frac{\omega_\mathrm{rf}^{2}}{4}\left(a_x + 2q_x \mathrm{cos}(\omega_\mathrm{rf} t)\right)
\end{equation}
where $a_x,\ q_x$ are Mathieu equation parameters \cite{leibfried_quantum_2003}, we obtain the solutions for $g_{\pm}(t),\ g_{0}(t)$ in terms of a single function $f$ as
\begin{equation}
\label{eqn_g_sol}
g_{-}(t)=\frac{|f|^2}{M}, \ g_{0}(t)=-\frac{|\dot{f}|^2}{2}, \ g_{+}(t)=M\left|\dot{f}\right|^2.
\end{equation}
The function $f_{1}=f$ and its conjugate $f_{2}=f^*$ are solutions to the Mathieu equation
\begin{equation}
\label{eqn_f_eqn}
\ddot{f_{i}}+\frac{\omega_\mathrm{rf}^{2}}{4}\left(a_x+2q_x \mathrm{cos}(\omega_\mathrm{rf} t)\right)f_{i}=0
\end{equation}
subject to the initial conditions $f(0)=1$ and $\dot{f}(0)=i\omega_x$ \cite{glauber_quantum}. It follows that $\omega_\mathrm{I}=\omega_x$, which is interpreted as the secular frequency of the trapped ion. The lowest order solution ($|a_x|,\ q_{x}^2 \ll 1$) is found to be
\begin{equation}
\label{eqn_f_sol}
f\approx e^{i\omega_x t}\frac{1+\frac{q_x}{2}\mathrm{cos}(\omega_\mathrm{rf}t)}{1+\frac{q_x}{2}}
\end{equation}
which can be substituted into Eq.~(\ref{eqn_g_sol}) to obtain
\begin{equation}
\label{eqn_g_minus_sol}
g_{-}(t)=\frac{1}{M}\left(\frac{1+\frac{q_x}{2}\mathrm{cos}(\omega_\mathrm{rf}t)}{1+\frac{q_x}{2}}\right)^2.
\end{equation}
Recovering $\hbar$, we obtain
\begin{equation}
\begin{gathered}
\label{eqn_apos_2}
x(t)=\left(\frac{1+\frac{q_x}{2}\mathrm{cos}(\omega_\mathrm{rf}t)}{1+\frac{q_x}{2}}\right) x_0 \left(e^{i\omega(t)}a^{\dag}+e^{-i\omega(t)}a\right) \\
+\left(\frac{1+\frac{q_x}{2}\mathrm{cos}(\omega_\mathrm{rf}t)}{1+\frac{q_x}{2}}\right)2x_0 \mathrm{Re}(\alpha(t))
\end{gathered}
\end{equation}
with
\begin{equation}
\label{eqn_omega_2}
\omega(t)=\omega_x \mathlarger{\mathlarger{\int}}_0^t dt'\left(\frac{1+\frac{q_x}{2}\mathrm{cos}(\omega_\mathrm{rf}t)}{1+\frac{q_x}{2}}\right)^2
\end{equation}
and 
\begin{multline}
\label{eqn_alpha_2}
\alpha(t)=e^{-i\omega(t)}\alpha(0) 
\\ +\frac{i}{\hbar}\mathlarger{\mathlarger{\int}}_{0}^{t}dt'e^{-i\left(\omega(t)-\omega(t')\right)}\left(\frac{1+\frac{q_x}{2}\mathrm{cos}(\omega_\mathrm{rf}t)}{1+\frac{q_x}{2}}\right)x_0F(t').
\end{multline}
Finally, we neglect the squeezing factor linked to intrinsic micromotion by making the approximation
\begin{equation}
\label{eqn_squeeze_approx}
\left(\frac{1+\frac{q_x}{2}\mathrm{cos}(\omega_\mathrm{rf}t)}{1+\frac{q_x}{2}}\right) \approx 1
\end{equation}
in Eqs. ~(\ref{eqn_apos_2}) and ~(\ref{eqn_omega_2}). This cannot be applied to the factor in the integral in Eq.~(\ref{eqn_alpha_2}) since it would amount to removing the effects of excess micromotion as well.

\nolinenumbers
%\nocite{*}
%\bibliographystyle{ieeetr}
\bibliography{draft1, draft2}

\end{document}